\documentclass[a4paper,12pt]{article}
\usepackage{epsfig}
\usepackage{amssymb}
\usepackage{cite}

\begin{document}

\begin{flushright}
  DCPT-08-60\\
  IPPP-08-30\\
  SHEP-08-16\\
May 22, 2008
\end{flushright}

\vfill

\begin{center}
\renewcommand{\thefootnote}{\fnsymbol{footnote}}

\textbf{\Large
The Scope of the 4$\tau$ Channel\\
in Higgs-strahlung and Vector Boson Fusion\\
for the NMSSM No-Lose Theorem at the LHC\\}

\vspace{10mm}

{\large
Alexander Belyaev$^{1,2}$, Stefan Hesselbach$^3$, Sami Lehti$^4$,  
Stefano Moretti$^{1}$,
Alexander Nikitenko$^{5,}$\footnote{~On leave from ITEP, Moscow, Russia.}, 
Claire H. Shepherd-Themistocleous$^2$}
\vspace{6mm}

{\itshape
$^1$School of Physics \& Astronomy, University of Southampton,
Highfield, Southampton SO17 1BJ, UK\\
$^2$Particle Physics Department, Rutherford Appleton Laboratory, Chilton,
Didcot, Oxon OX11 0QX, UK\\
$^3$IPPP, University of Durham, South Road, Durham DH1 3LE, UK\\
$^4$Helsinki Institute of Physics, P.O. Box 64, 
00014 Helsinki, Finland\\
$^5$Blackett Laboratory, Physics Department,
Imperial College, %Prince Consort Road,
London SW7 2AZ, UK}

\end{center}

\vfill

\begin{abstract}
\noindent
We {study}  the potential of the $h_1\to a_1a_1\to 4\tau$ 
{signal from}  the lightest scalar ($h_1$) and pseudoscalar ($a_1$)
Higgs bosons {to cover the parameter space} 
of the Next-to-Minimal Supersymmetric Standard Model (NMSSM) at the Large 
Hadron Collider (LHC). We exploit {a  $2\mu+2$jets signature
from four $\tau$'s decays} (accompanied by missing transverse energy), 
resorting to both Higgs-strahlung (HS), by
triggering on leptonic $W^\pm$ decays,
and Vector Boson Fusion (VBF), by
triggering on two same sign non-isolated muons.  
\end{abstract}

\vfill
\newpage
\setcounter{footnote}{0}

%%%%%%%%%%%%%%%%%%%%%%%%%%%%%%%%%%%
\section{Introduction}

The
NMSSM
\cite{Nilles:1982dy,Frere:1983ag,Ellis:1988er,Drees:1988fc,Ellwanger:1993hn,Ellwanger:1993xa,Elliott:1993bs,Pandita:1993tg,Ellwanger:1995ru,King:1995vk,Franke:1995tc,Ellwanger:1996gw,Miller:2003ay},
which contains a singlet/singlino superfield in addition to
the particle content of the Minimal Supersymmetric Standard Model (MSSM),
has various advantages with respect to the MSSM {such as} the solution of the $\mu$
problem and the smaller fine-tuning%
\footnote{Alternative formulations of the MSSM with an additional
singlet/singlino superfield --- known
as the Minimal Non-minimal Supersymmetric Standard Model (MNSSM) and
new Minimally-extended Supersymmetric Standard Model or
nearly-Minimal Supersymmetric Standard Model (nMSSM) --- also
exist \cite{Panagiotakopoulos:1999ah,Panagiotakopoulos:2000wp,Dedes:2000jp,Hugonie:2001ib}.}
\cite{Djouadi:2008uw,Djouadi:2008yj,cNMSSM-benchmarks}.
The additional singlet field results in one new CP-even and one new CP-odd
state in 
the Higgs sector which consists of three CP-even mass eigenstates ($h_{1,2,3}$)
and two CP-odd states ($a_{1,2}$) in the NMSSM.
{In contrast  to the MSSM,
it is 
not certain that even at least one Higgs boson can be found at the 
CERN Large Hadron Collider (LHC) within the NMSSM.
This problem can be created by  the additional singlet states in the Higgs sector
leading to invisible decay channels 
of the SM-like Higgs states.}
{Of particular relevance for the Higgs search at the LHC in the NMSSM
are  the decays of the lightest CP-even Higgs into lighter CP-odd
states, $h_1\to a_1a_1$. This channel
has been claimed to be the only means to establish
a no-lose theorem for the NMSSM
\cite{Gunion:1996fb,Ellwanger:2001iw,Azuelos:2002qw,Ellwanger:2003jt,%
Ellwanger:2004gz,Miller:2004uh,Assamagan:2004mu,Weiglein:2004hn,%
Moretti:2006hq,Accomando:2006ga,Carena:2007jk}, 
at least in parameter regions where the Supersymmetric (SUSY) partners
of ordinary Standard 
Model (SM) objects are made suitable heavy.} Also notice that
the $h_1$ state could well be below the LEP limit
on the SM Higgs mass, of 114 GeV, albeit with weakened couplings to ordinary
matter.

In this letter we will focus on NMSSM parameter regions with light $a_1$ states
(light $a_1$ scenario)
with $M_{a_1}< 2 m_b$ where the decay $a_1 \to b \bar{b}$
is not possible.
In these parameter regions with large BR($a_1 \to \tau^+\tau^-$) 
the scope of $h_1\to a_1a_1$ decays into $jj\tau^+\tau^-$ pairs
(where $j$ represents a jet of either heavy or light flavour and where
the $\tau$'s decay leptonically) has been found to be rather questionable
\cite{Baffioni:2004gdr}. Hence, here we 
investigate the scope of the $4\tau$ channel, wherein two $\tau$'s are searched
for in their muonic decays, while the other two are selected via their
hadronic ones. {In doing so, 
we exploit both the HS and VBF  production channels.}

The plan of the paper is as follows.
In section 2 the results of a detailed parameter scan are discussed identifying
the parameter regions with light $a_1$ and possibly light $h_1$ with
$M_{h_1}<114$ GeV. 
In section 3 the $h_1$ production and decay processes are simulated
with the PYTHIA event generator 
\cite{Sjostrand:2006za}
and
the prospects for the determination of the NMSSM Higgs bosons in
the channel  $h_1\to a_1a_1 \to 4 \tau$ are discussed. We then conclude
in section 4.

\section{The low-energy NMSSM parameter space for the light $a_1$ scenario}

Here we investigate the NMSSM {parameter space for $M_{a_1}< 2 m_b$},
with particular interest
in the cases where the aforementioned $h_1\to a_1a_1\to 4\tau$ decays
may be visible at the LHC {from HS and/or VBF 
production processes.}
The NMSSM Higgs at the Electro-Weak (EW) scale is uniquely defined
by fourteen parameters:
the ratio  of the doublet Higgs vacuum expectation values (VEVs) $\tan\beta$,
the trilinear couplings in the superpotential $\lambda$ and $\kappa$, the
corresponding soft SUSY breaking parameters $A_\lambda$ and $A_\kappa$,
the effective $\mu$ parameter $\mu = \lambda \langle S \rangle$
(where $\langle S \rangle$ denotes the VEV of the singlet Higgs field),
the gaugino mass parameters $M_1$, $M_2$ and $M_3$,
the squark and slepton trilinear couplings
$A_{t}$,  $A_{b}$ and  $A_\tau$,
and the squark and slepton mass parameters $M_{f_L}$ and $M_{f_R}$.
In the following we will establish the NMSSM parameter regions, defined
in terms of the above inputs, that survive present theoretical and 
experimental constraints. 

\subsection{Full NMSSM parameter scan}

First we perform a `wide' scan over the full NMSSM parameter space,
where the numerical values over which the parameters have
been varied are:
$$
-1000~{\rm{GeV}}<A_\kappa<100~{\rm{GeV}}, ~~
-5~{\rm{TeV}}<A_\lambda<5~{\rm{TeV}}, ~~
100~{\rm{GeV}}<\mu<1000~{\rm{GeV}},
$$
\begin{equation}
10^{-5}<\lambda,\kappa<0.7, ~~
  1.5<\tan\beta<50
\label{par-space}
\end{equation}
while the remaining parameters (entering the Higgs sector at
loop-level) were fixed at
\begin{equation}
M_1/M_2/M_3=150/300/1000~{\rm{GeV}},~~
A_{t}=A_{b}=A_\tau=2.5~{\rm{TeV}},~~
M_{f_L}=M_{f_R} =1~{\rm{TeV}},
\label{par-fixed}
\end{equation}
resulting in a heavy spectrum of the SUSY partners of
$\mathcal{O}=1$~TeV (except for the U(1) and SU(2) gauginos).
This scan (as well
as all {the others} in the remainder of this letter) has been performed by using the
NMSSMTools 
package \cite{Ellwanger:2004xm,Ellwanger:2005dv,Domingo:2007dx},
which calculates NMSSM spectra (masses, couplings and decay rates) and takes
into account 
experimental inputs including LEP limits, $B$-physics bounds as well
as constraints on the relic density of
cold dark matter (DM) stemming from the results obtained by the Wilkinson
Microwave Anisotropy Probe (WMAP)~\cite{wmap}.

\begin{figure}[!t]
  \includegraphics[width=0.38\textwidth]{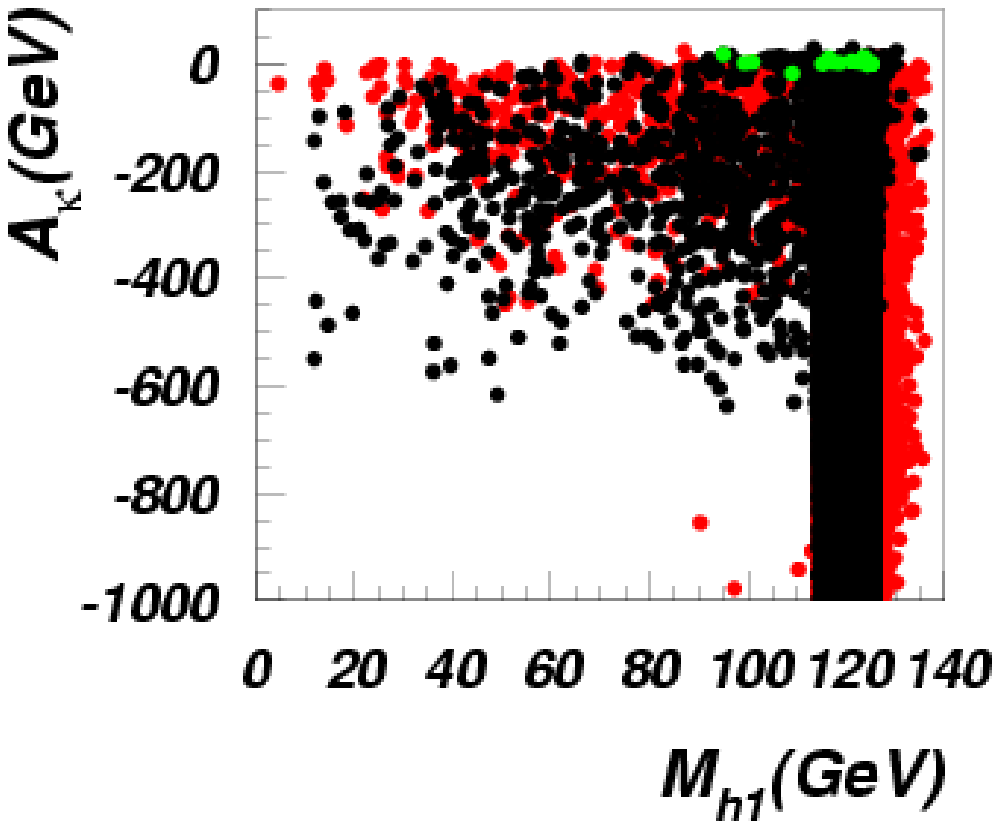}\hspace*{-0.9cm}
  \includegraphics[width=0.38\textwidth]{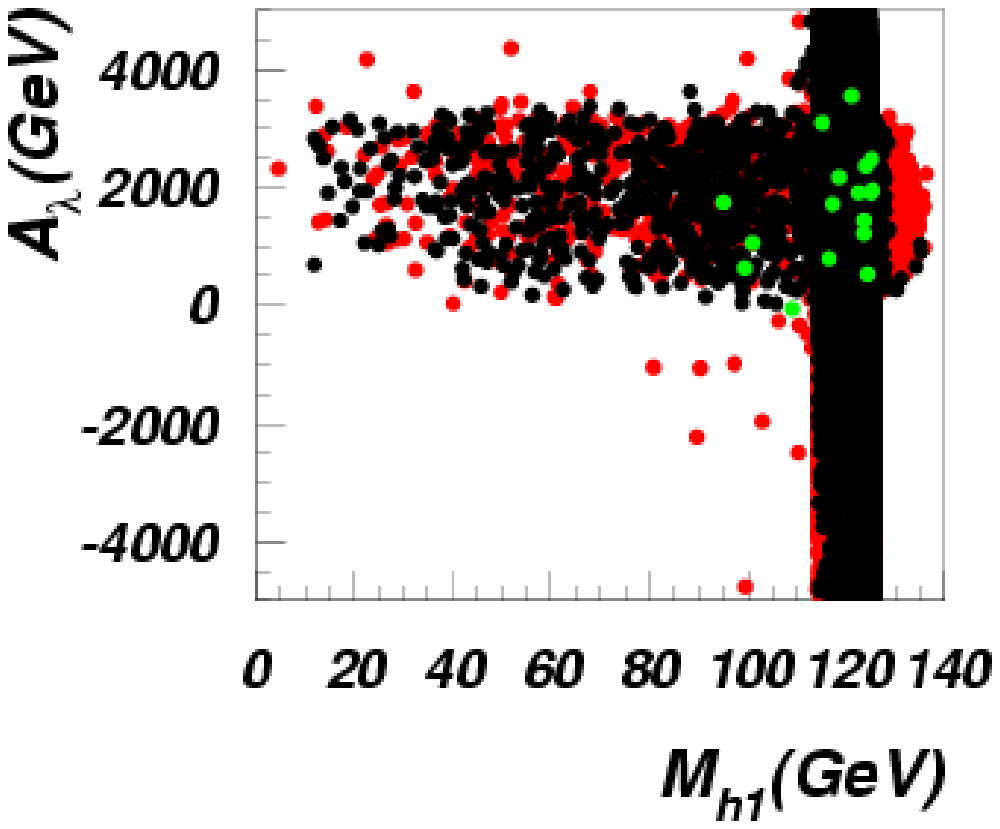}\hspace*{-0.9cm}
  \includegraphics[width=0.38\textwidth]{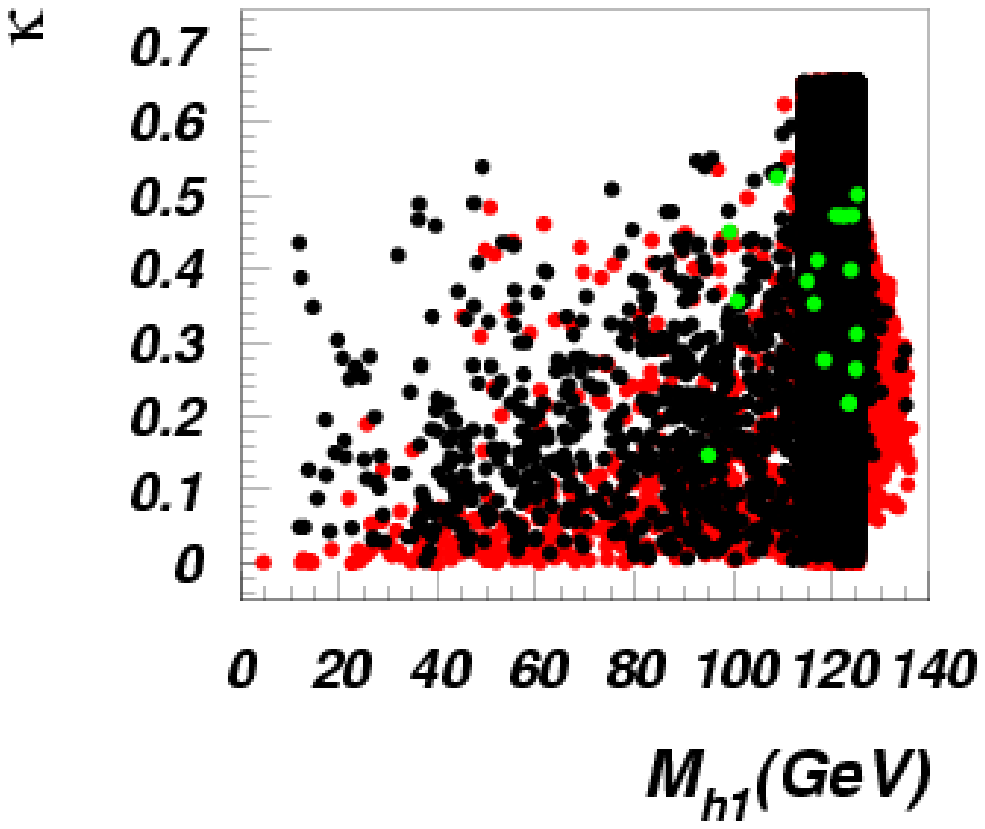} \hspace*{-0.9cm}
\vskip -2.5cm
\hspace*{0.33\textwidth}\hspace*{-5.5cm}{\bf (a)}
\hspace*{0.33\textwidth}\hspace*{-0.6cm}{\bf (b)}
\hspace*{0.33\textwidth}\hspace*{-0.6cm}{\bf (c)}
\vskip  1.5cm
  \includegraphics[width=0.38\textwidth]{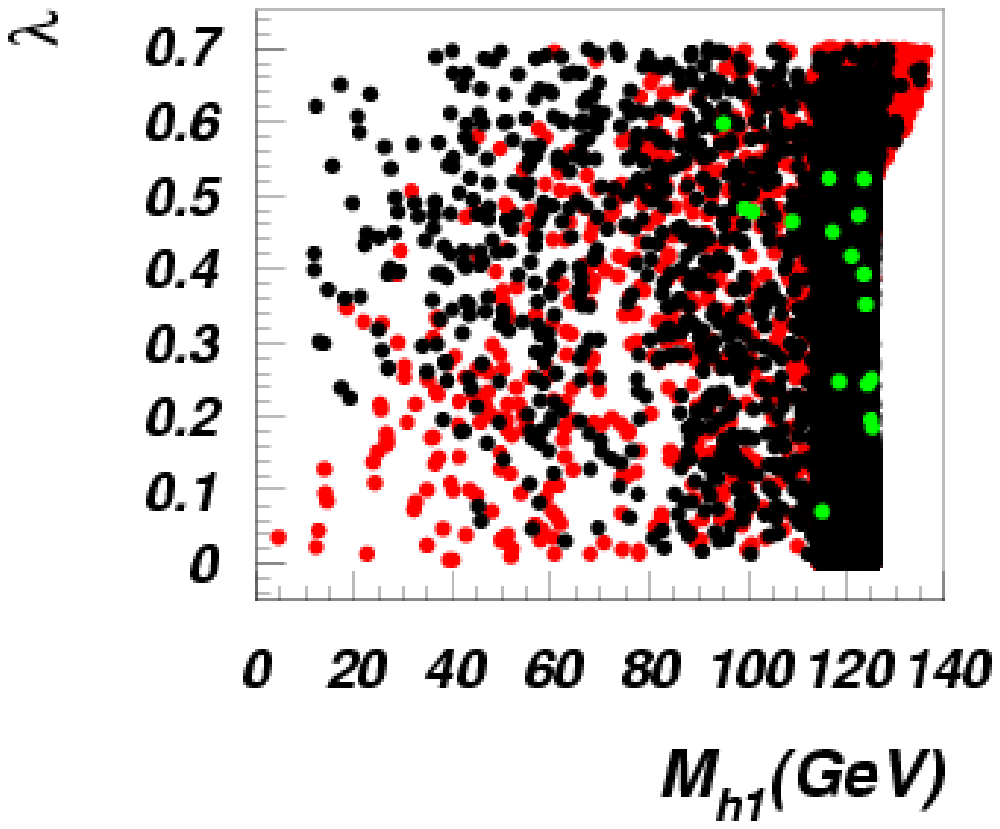}\hspace*{-0.9cm}
  \includegraphics[width=0.38\textwidth]{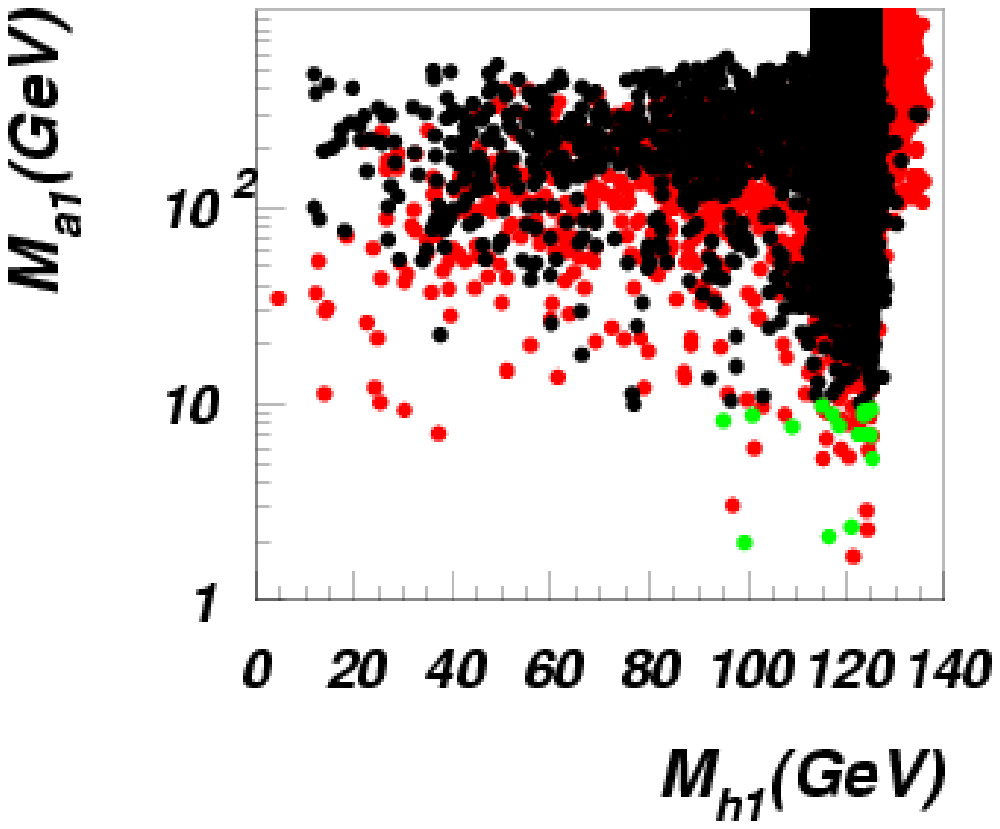}\hspace*{-0.9cm}
  \includegraphics[width=0.38\textwidth]{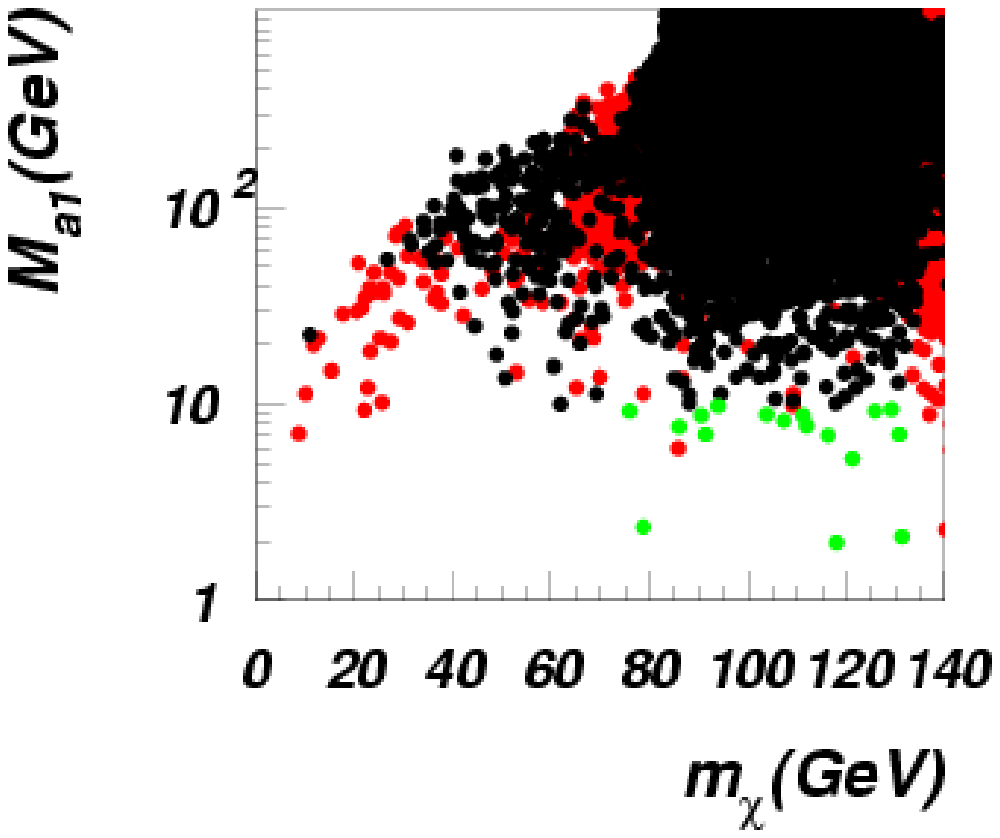}\hspace*{-0.9cm}
\vskip -2.5cm
\hspace*{0.33\textwidth}\hspace*{-5.5cm}{\bf (d)}
\hspace*{0.33\textwidth}\hspace*{-0.6cm}{\bf (e)}
\hspace*{0.33\textwidth}\hspace*{-0.6cm}{\bf (f)}
\vskip  2cm
\caption{Results of the `wide' scan over the full NMSSM parameter
 	 space, Eq.~(\ref{par-space}), mapped onto the planes:
 	 (a) [$A_\kappa, M_{h_1}$],
 	 (b) [$A_\lambda, M_{h_1}$],
 	 (c) [$\kappa, M_{h_1}$],
 	 (d) [$\lambda, M_{h_1}$],
 	 (e) [$M_{a_1}, M_{h_1}$],
         (f) [$M_{a_1}, m_{\chi^0_1}$].
 Colour code:
 red    -- all constraints are satisfied but relic density (above WMAP constraint: $\Omega h^2>0.11$);
 black  -- all constraints are satisfied, $M_{a_1}>10$~GeV;
 green -- all constraints are satisfied, $M_{a_1}<10$~GeV. 
\label{fig:scan-wide}}
\end{figure}

In Fig.~\ref{fig:scan-wide} we present the results of this scan in
form of scatter plots,
corresponding to $10^7$ generated points, where about $2.7\times 10^5$
points survive all constraints but the DM constraints (red points) and
about $7.6\times 10^4$ points survive all constraints (black and green
points). The bands between
$115~\mathrm{GeV} \lesssim M_{h_1} \lesssim 130~\mathrm{GeV}$
appear because in a majority of the generated points the $h_1$ has
SM-like properties where the lower bound of the band is given by the
LEP Higgs limit and the upper bound by the theoretical upper limit
of the $h_1$ mass in the NMSSM.
For the points with $M_{h_1} \lesssim 115~\mathrm{GeV}$ the $h_1$
couplings to SM gauge bosons are suppressed, hence the LEP Higgs limit
is not applicable. 
Though only 17 points with $M_{a_1}<10$~GeV survive (green points),
one can see from Fig.~\ref{fig:scan-wide}(a) 
that small $|A_\kappa|$'s are preferred
while 
Fig.~\ref{fig:scan-wide}(b)  indicates
the preference for large positive $A_\lambda$.

\subsection{Scan for narrowed $A_\kappa$ range}

\begin{figure}[!t]
  \includegraphics[width=0.38\textwidth]{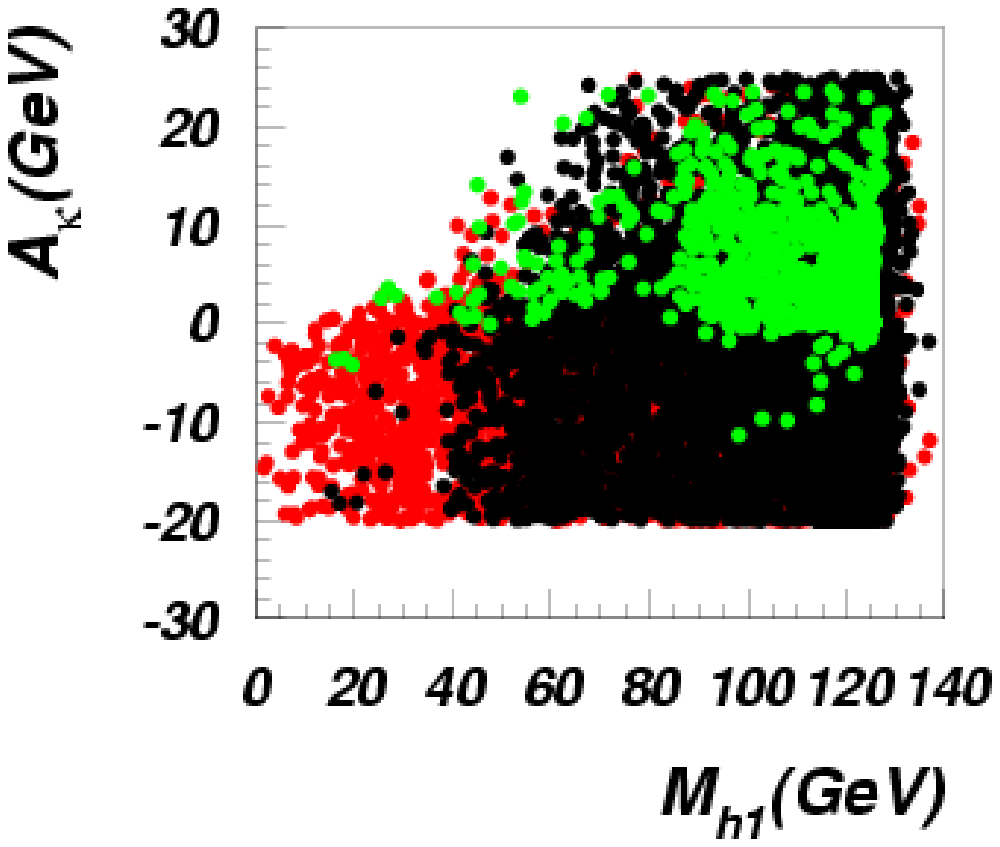}\hspace*{-0.9cm}
  \includegraphics[width=0.38\textwidth]{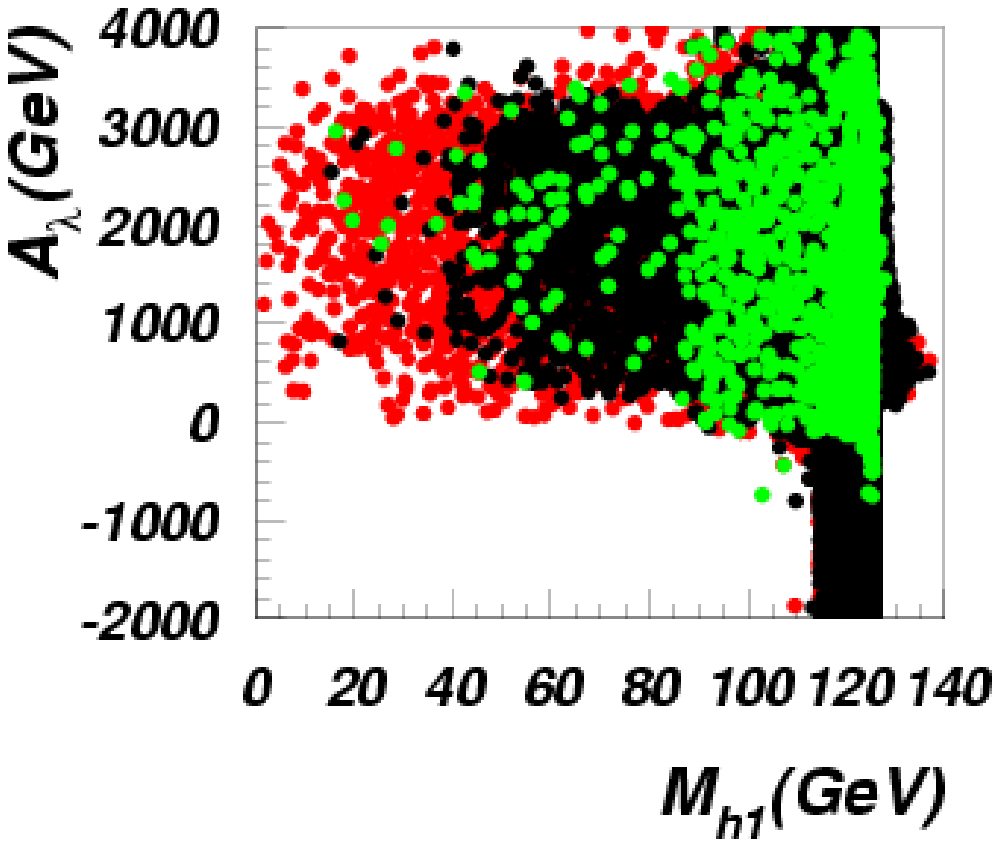}\hspace*{-0.9cm}
  \includegraphics[width=0.38\textwidth]{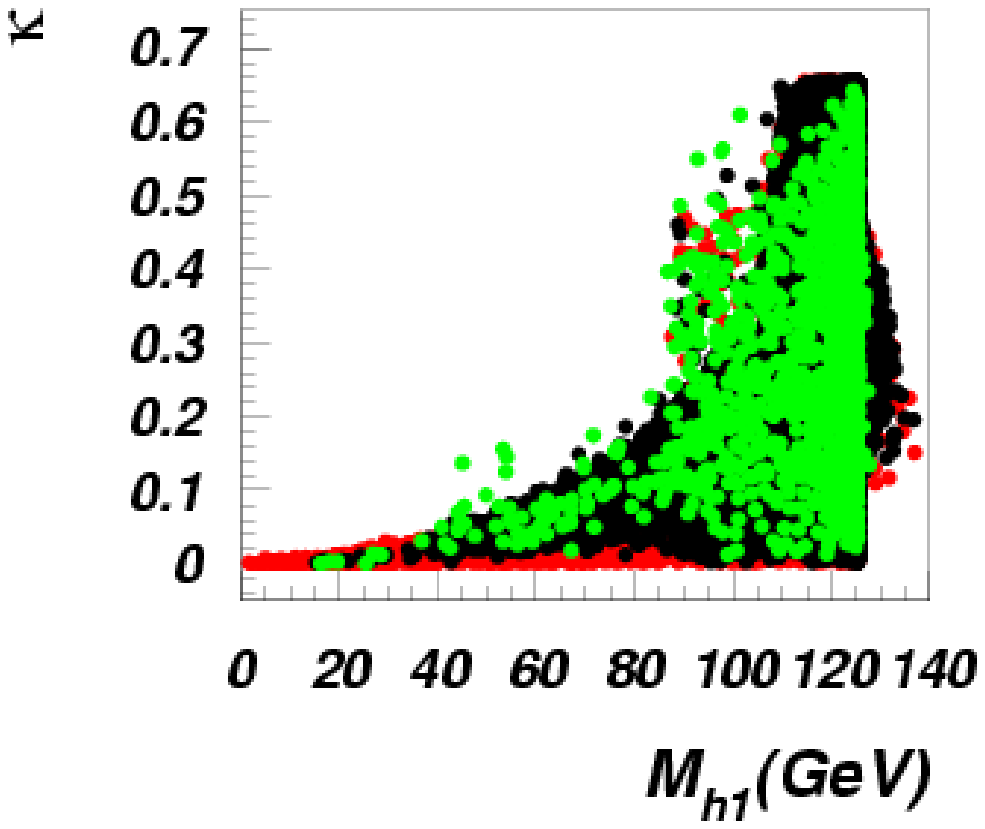} \hspace*{-0.9cm}
\vskip -2.5cm
\hspace*{0.33\textwidth}\hspace*{-5.5cm}{\bf (a)}
\hspace*{0.33\textwidth}\hspace*{-0.6cm}{\bf (b)}
\hspace*{0.33\textwidth}\hspace*{-0.6cm}{\bf (c)}
\vskip  1.5cm
  \includegraphics[width=0.38\textwidth]{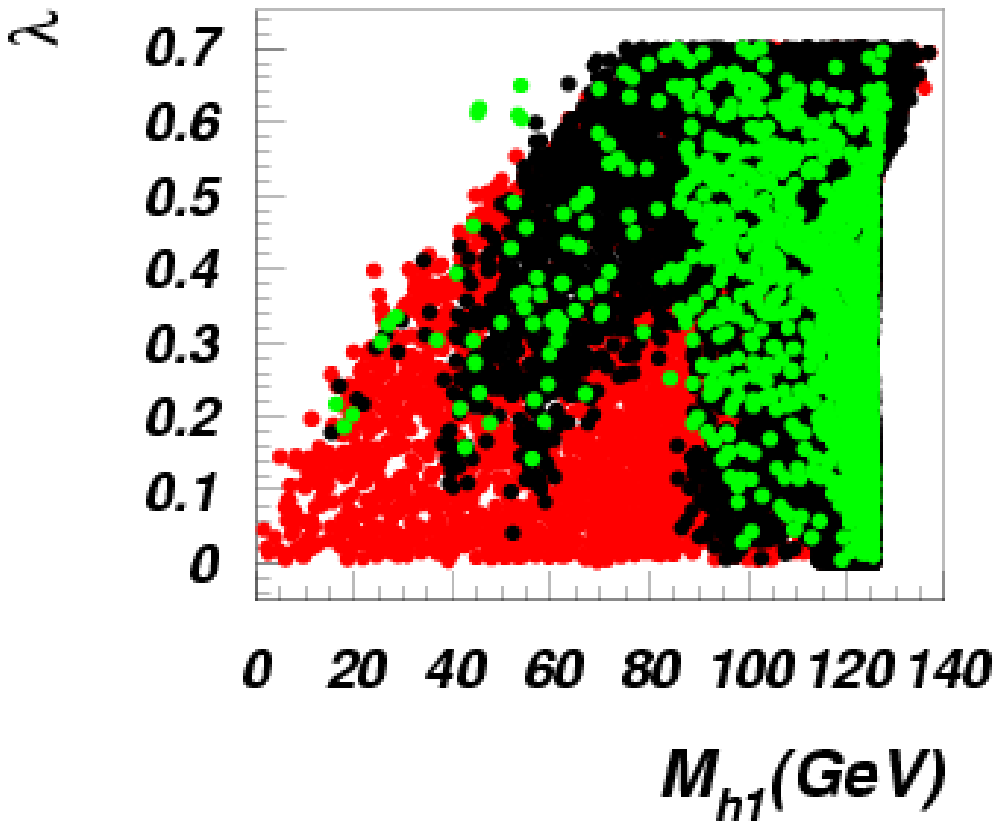}\hspace*{-0.9cm}
  \includegraphics[width=0.38\textwidth]{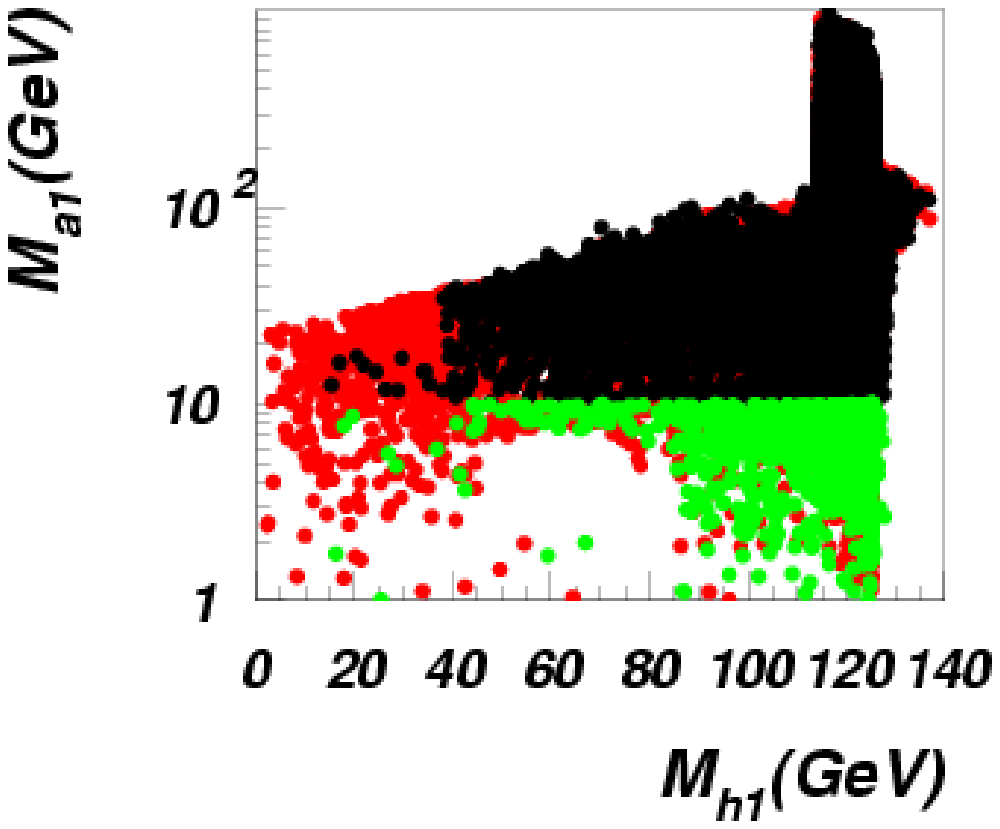}\hspace*{-0.9cm}
  \includegraphics[width=0.38\textwidth]{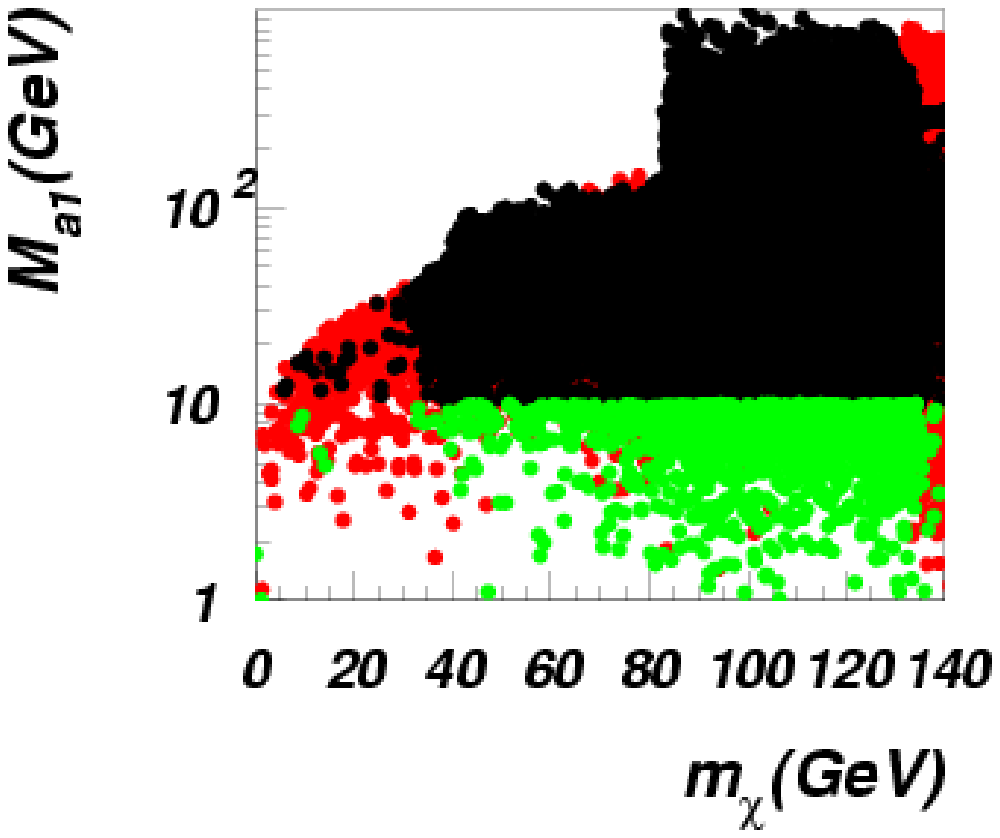}\hspace*{-0.9cm}
\vskip -2.5cm
\hspace*{0.33\textwidth}\hspace*{-5.5cm}{\bf (d)}
\hspace*{0.33\textwidth}\hspace*{-0.6cm}{\bf (e)}
\hspace*{0.33\textwidth}\hspace*{-0.6cm}{\bf (f)}
\vskip  2cm
\caption{Results of the scan over the `narrowed' NMSSM parameter
  space, i.e., analogous to  Fig.~\ref{fig:scan-wide}
but for $-20~{\rm{GeV}}<A_\kappa<25~{\rm{GeV}}$,
$-2~{\rm{TeV}}<A_\lambda<4~{\rm{TeV}}$,
$100~{\rm{GeV}}<\mu<300~{\rm{GeV}}$.
The individual plots and the colour code are the same as in Fig.~\ref{fig:scan-wide}.
\label{fig:scan-narrow}}
\end{figure}

The results of Fig.~\ref{fig:scan-wide}
({especially} the preference for small $A_\kappa$'s) motivated us
to `narrow' the range of the parameters, by scanning it over the intervals
\begin{equation}
-20~{\rm{GeV}}<A_\kappa<25~{\rm{GeV}},~~
-2~{\rm{TeV}}<A_\lambda<4~{\rm{TeV}}, ~~
100~{\rm{GeV}}<\mu<300~{\rm{GeV}},
  \label{cut:ak_narrow}
\end{equation}
and the rest of the parameters as in Eq.~(\ref{par-space}).
Fig.~\ref{fig:scan-narrow},
based on $10^7$ generated points resulting in $2.9 \times 10^5$ points
surviving all but the DM constraints (red points) and
$4.0 \times 10^5$ points surviving all constraints (black and green points),
shows that this is precisely the region
where a large portion of the NMSSM parameter points with
$M_{a_1}<10$~GeV (here $3.3\times 10^3$ green points) are found.
Now we can also see certain {correlations and other 
interesting features} onsetting in the $M_{a_1}<10$~GeV
region:
\begin{enumerate}
\item values of $A_\lambda>0$ are preferred,
    see Fig.~\ref{fig:scan-narrow}(b);
\item {for points with low $M_{h_1}$ small values of $|A_\kappa|$
   (Fig.~\ref{fig:scan-narrow}(a)) as well as 
   small values of $\kappa$ 
   (Fig.~\ref{fig:scan-narrow}(c)) are preferred};
\item {in the $M_{a_1}<10$~GeV region of our interest,
      $M_{h_1}$ as low as 20 GeV can appear
      (Fig.~\ref{fig:scan-narrow}(e)).}
\end{enumerate}

\subsection{Final scan for the light $a_1$ scenario}

\begin{figure}[!t]
  \includegraphics[width=0.38\textwidth]{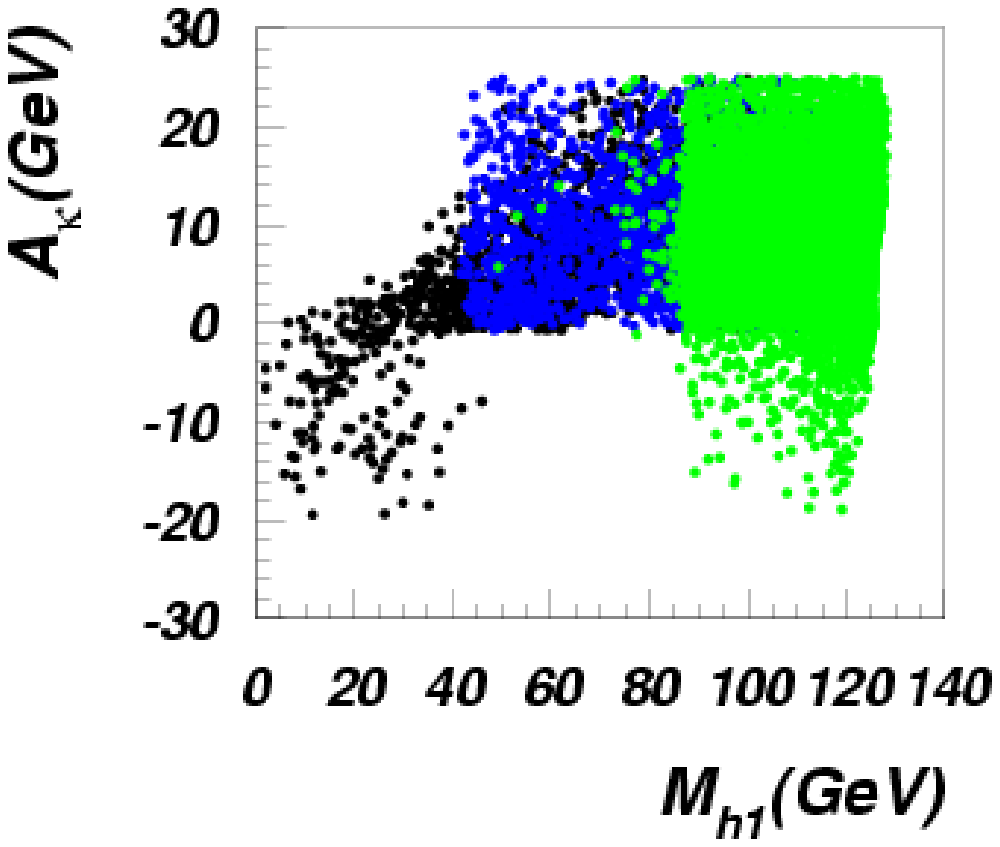}\hspace*{-0.9cm}
  \includegraphics[width=0.38\textwidth]{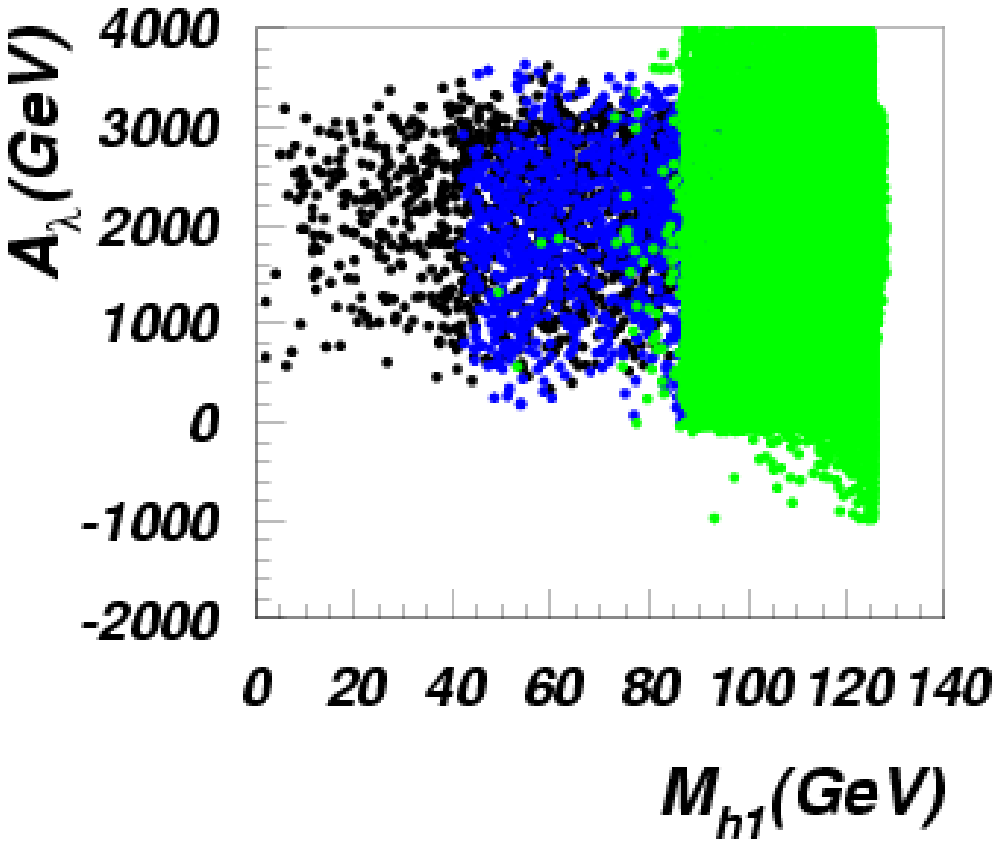}\hspace*{-0.9cm}
  \includegraphics[width=0.38\textwidth]{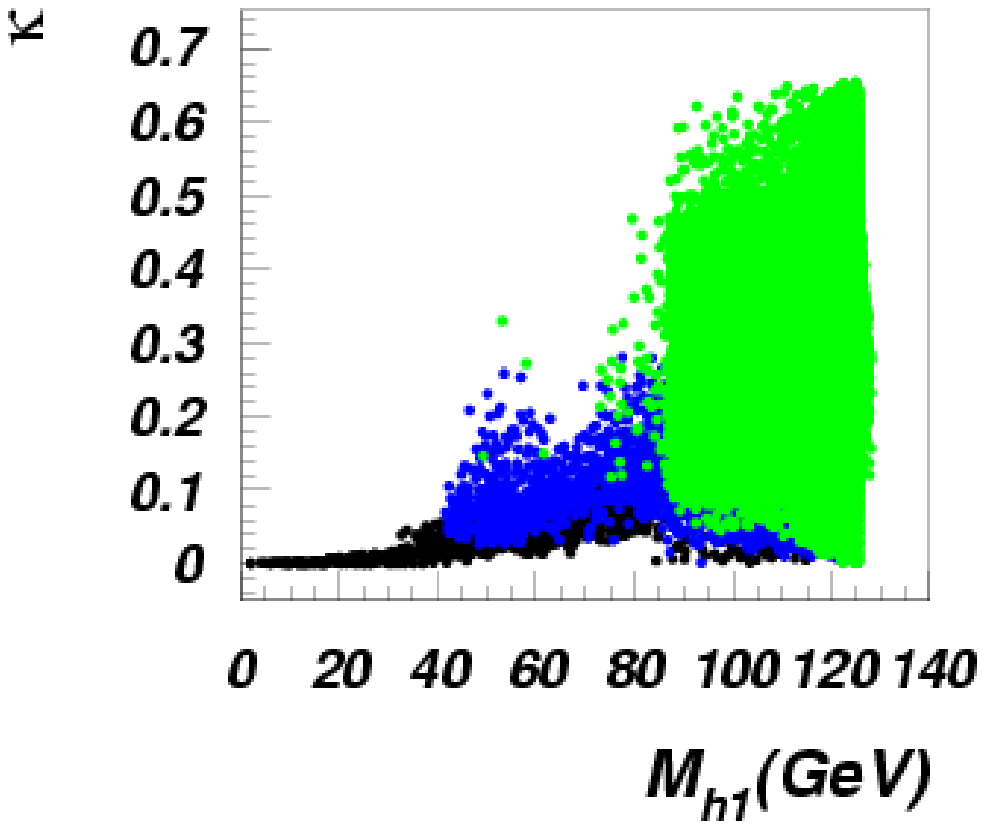} \hspace*{-0.9cm}
\vskip -2.5cm
\hspace*{0.33\textwidth}\hspace*{-5.5cm}{\bf (a)}
\hspace*{0.33\textwidth}\hspace*{-0.6cm}{\bf (b)}
\hspace*{0.33\textwidth}\hspace*{-0.6cm}{\bf (c)}
\vskip  1.5cm
  \includegraphics[width=0.38\textwidth]{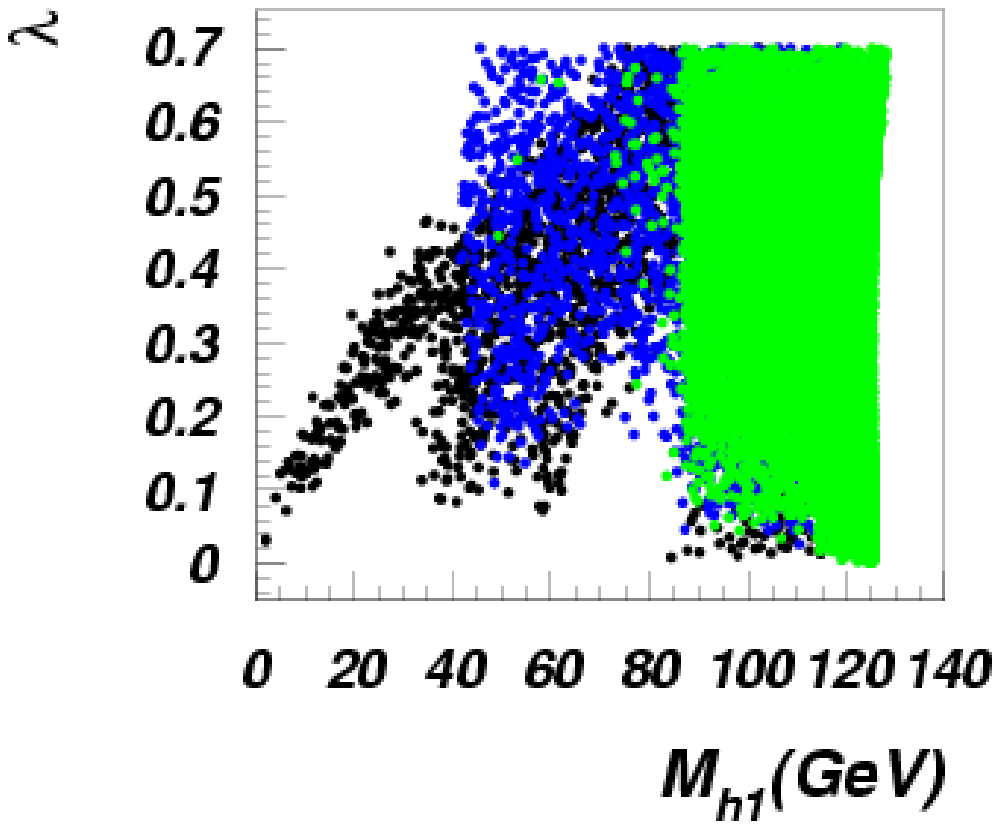}\hspace*{-0.9cm}
  \includegraphics[width=0.38\textwidth]{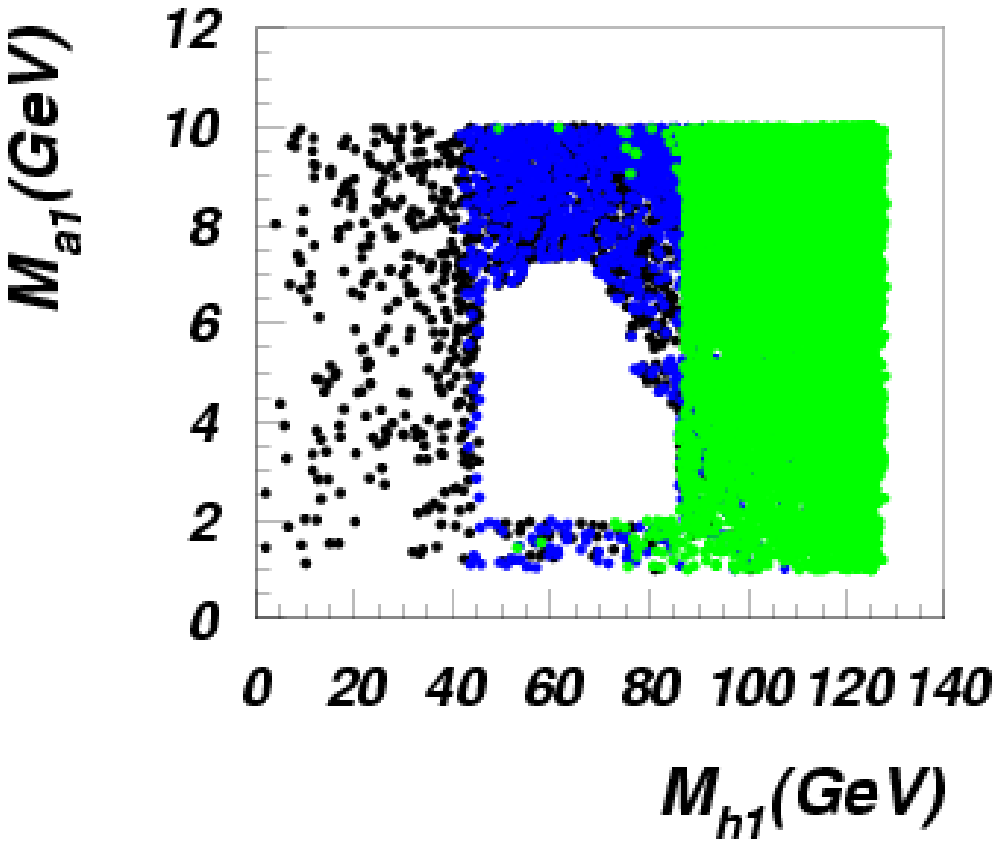}\hspace*{-0.9cm}
  \includegraphics[width=0.38\textwidth]{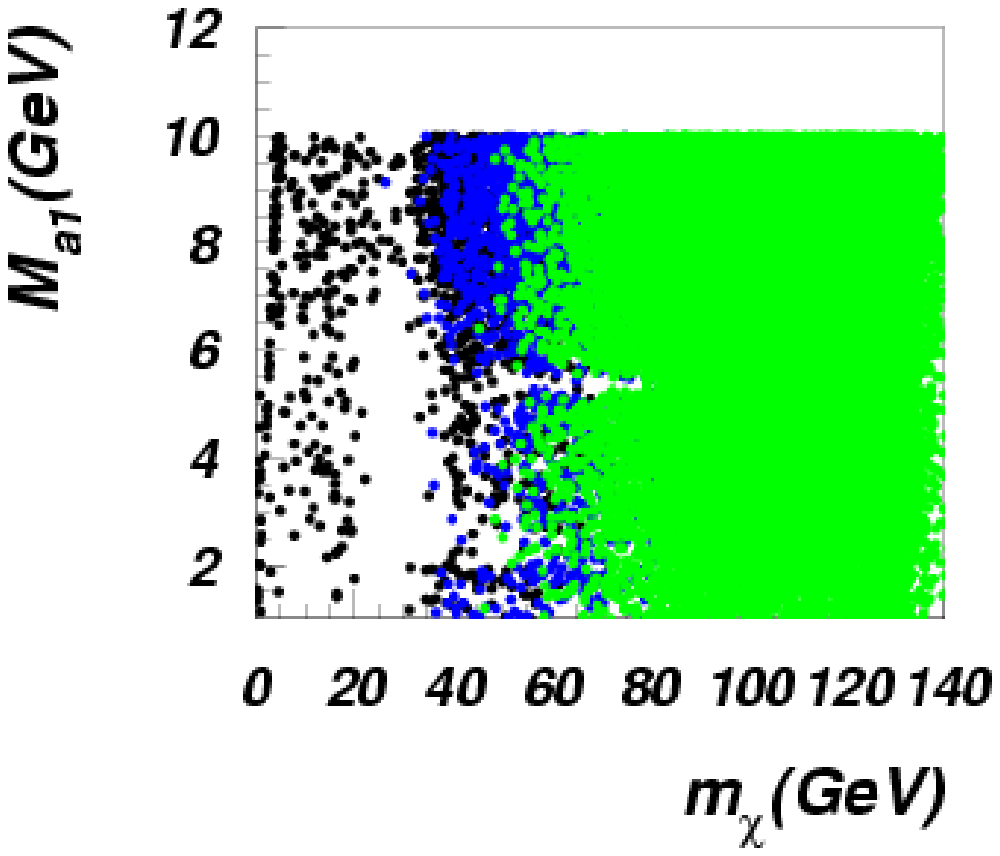}\hspace*{-0.9cm}
\vskip -2.5cm
\hspace*{0.33\textwidth}\hspace*{-5.5cm}{\bf (d)}
\hspace*{0.33\textwidth}\hspace*{-0.6cm}{\bf (e)}
\hspace*{0.33\textwidth}\hspace*{-0.6cm}{\bf (f)}
\vskip  2cm
\caption{\label{fig:scan-final}
Results of the `final' NMSSM parameter scan, i.e., with $M_{a_1}< 10$
GeV and with Eq.~(\ref{cut:ak_narrow})
enforced. The black, blue and green colours  indicate  the cases
 $R_{ZZh}<0.1$,
 $0.1<R_{ZZh}<0.5$ and
 $R_{ZZh}>0.5$, respectively (where $R_{ZZh}$ is defined in the text).
The individual plots are the same as in Fig.~\ref{fig:scan-wide}.}
\end{figure}

We have then performed a `final' scan over the
NMSSM parameter space by requiring $M_{a_1}< 10$~GeV additionally to
Eq.~(\ref{cut:ak_narrow}).
{At this stage, 
we recast the NMSSM parameter space in terms 
of the quantities entering $h_1$ HS and VBF production and  $h_1\to a_1a_1\to 4\tau$
decays.}
The results of this scan 
($10^8$ generated points, resulting in $3.9\times 10^4$ points with
$M_{a_1}< 10$~GeV surviving all constraints)
are shown in Fig.~\ref{fig:scan-final},
where the colours were chosen to indicate the measure of decoupling of the
lightest CP-even Higgs boson, $h_1$, from the SM limit (denoted simply by $H$).
{This measure of decoupling
was defined through  the ratio of the coupling strength (squared) of the $ZZh_1$ vertex in
the NMSSM relative to the SM case (in fact,
this is the same for couplings to $W^\pm$ gauge bosons):
$R_{ZZh}=\left(g_{ZZh_1}^{\rm{NMSSM}}/g_{{ZZH}}^{\rm{SM}}\right)^2$).}
One should notice that both HS, $pp\to V h_1$, and VBF,
$pp\to jjV^*V^*\to jjh_1$, rates ($V=Z,W^\pm$) are directly proportional to
$R_{ZZh}$ 
and are suppressed in the non-decoupling regime
whenever $R_{ZZh}$ is essentially smaller then unity.

From Figs.~\ref{fig:scan-final} and \ref{final-scan1}
 one can see the following important features of
the $M_{a_1}< 2m_b\approx10$ GeV  scenario:
\begin{enumerate}

\item The lighter the Higgs $h_1$ the more significant  
   should be the NMSSM deviations from the SM case, e.g.,
   for any $M_{h_1}<50$~GeV any $R_{ZZh}$ is limited  to be
   $<0.5$, as dictated by LEP 
   constraints~\cite{Barate:2003sz,unknown:2006cr}
   (this correlation is illustrated in a more clear way  in
   Fig.~\ref{final-scan1}(a), presenting the $R_{ZZh}$ versus $M_{h_1}$
   plane,
   which exhibits the typical pattern of the LEP Higgs exclusion 
   curve~\cite{unknown:2006cr}).

\item In the $M_{h_1}<40$ GeV region $A_\lambda$ is always positive
  (Fig.~\ref{fig:scan-final}(b)), $\lambda<0.45$
  (Fig.~\ref{fig:scan-final}(d)) {while $\kappa < 0.1$}
  (Figs.~\ref{fig:scan-final}(c), \ref{final-scan1}(b)), which
  corresponds to an approximate Peccei-Quinn (PQ) symmetry
  ($\kappa \to 0$ and $\kappa A_\kappa \to 0$) 
  \cite{Miller:2003ay,Miller:2005qua}.
  However, for $M_{h_1}\gtrsim M_Z$ the whole range
  $0< \kappa < \kappa_\mathrm{max} \sim 0.7$ is populated. 
  Hence an approximate PQ symmetry is not necessary for the
  $M_{a_1}< 2m_b\approx10$ GeV  {and  $M_{h_1}\gtrsim M_Z$} scenario while
  some fine-tuning ensuring $|A_\kappa| \ll |A_\lambda|$ is present.

\item 
 One should  further notice the correlation between the singlet
 nature of the $h_1$ and the singlino component of the
 lightest neutralino {as well as  the correlation between} their masses,
 which are visualised in
 Fig.~\ref{final-scan1}(c) and (d), respectively.
 These correlations take place to satisfy WMAP constraints
 on relic density. In fact, 
 {in the lower-left corner Fig.~\ref{final-scan1}(d) 
 we can clearly  observe $\chi^0_1\chi^0_1\to h_1$ the
 annihilation corridor for  $M_{h_1}<80$~GeV,
 along which $2m_{\chi^0_1} \simeq M_{h_1}$
 that allows one to lower down the neutralino relic density
 to a level consistent with experimental constraints.
 In this corridor  $\chi^0_1$ and $h_1$ have significant
 singlino and singlet components, respectively.
 At the very bottom of  Fig.~\ref{final-scan1}(d) 
 we can further see another DM motivated region
 with very light ($<5$ GeV) neutralino
 where  $2m_{\chi^0_1}  \simeq M_{a_1}$
 provides the $\chi^0_1\chi^0_1\to a_1$ annihilation corridor.
 This region is also represented  in Fig.~\ref{fig:scan-final}(f)
 as a border-line at the left edge of the parameter space.
These two representative  DM motivated regions appear because
the NMSSM structure
requires $h_1$ to be a singlet
and $\chi^0_1$ to be a singlino for $M_{a_1}<10$~GeV and  low $M_{h_1}$.
}

\begin{figure}[!t]
  \includegraphics[width=0.38\textwidth]{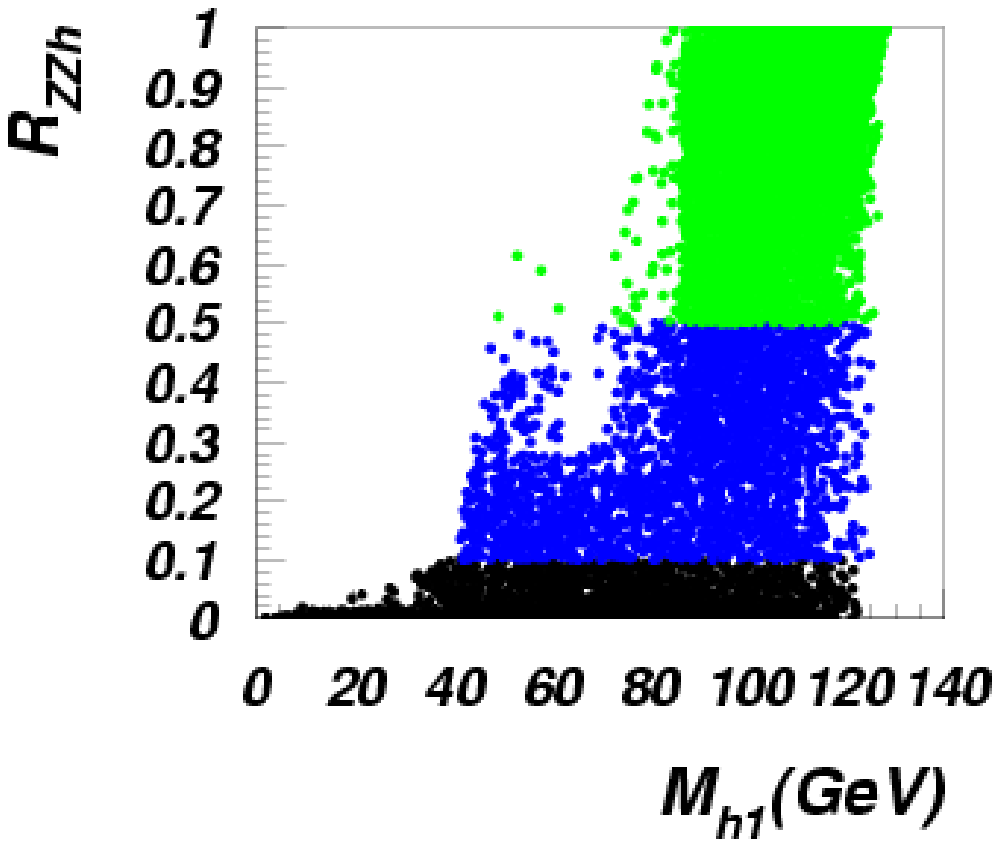}\hspace*{-0.9cm}
  \includegraphics[width=0.38\textwidth]{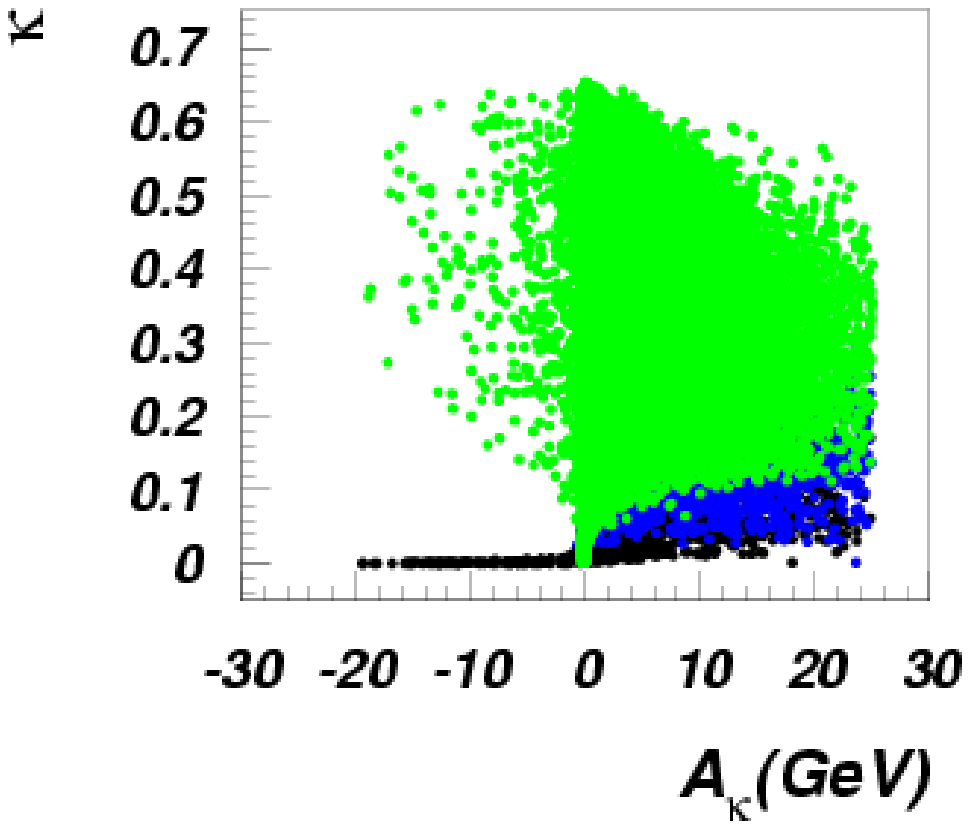}\hspace*{-0.9cm}
  \includegraphics[width=0.38\textwidth]{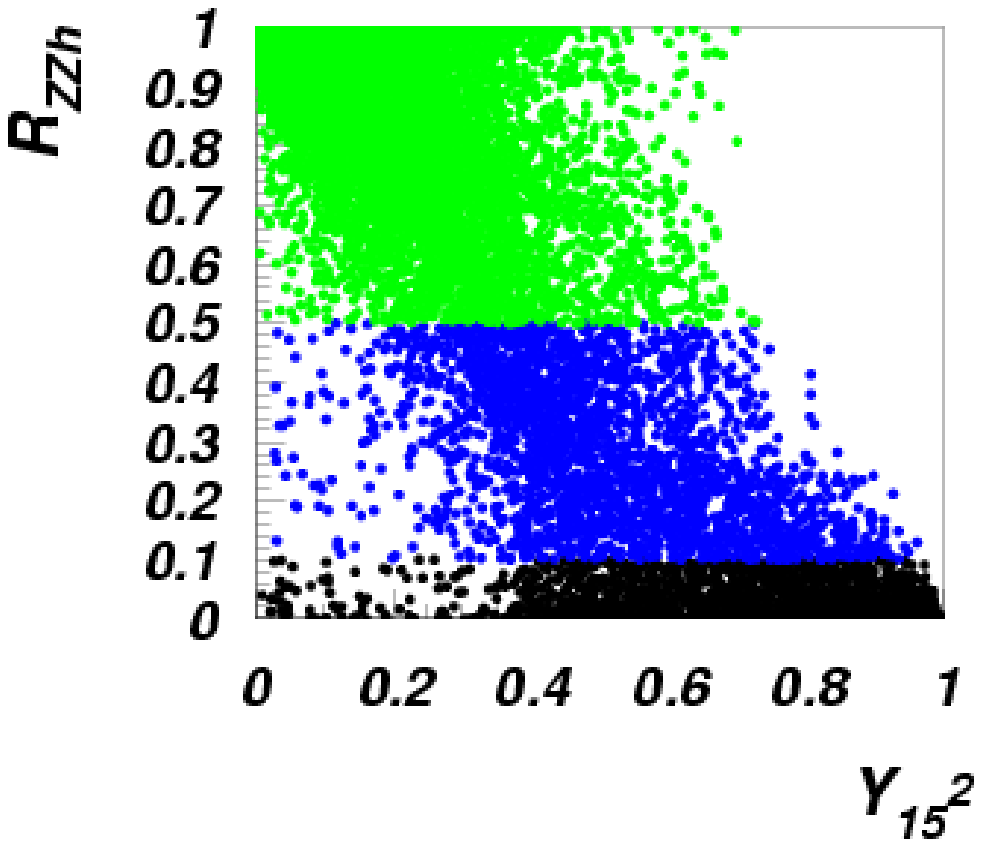}\hspace*{-0.9cm}
\vskip -2.5cm
\hspace*{0.33\textwidth}\hspace*{-5.5cm}{\bf (a)}
\hspace*{0.33\textwidth}\hspace*{-0.6cm}{\bf (b)}
\hspace*{0.33\textwidth}\hspace*{-0.6cm}{\bf (c)}
\vskip  1.5cm
  \includegraphics[width=0.38\textwidth]{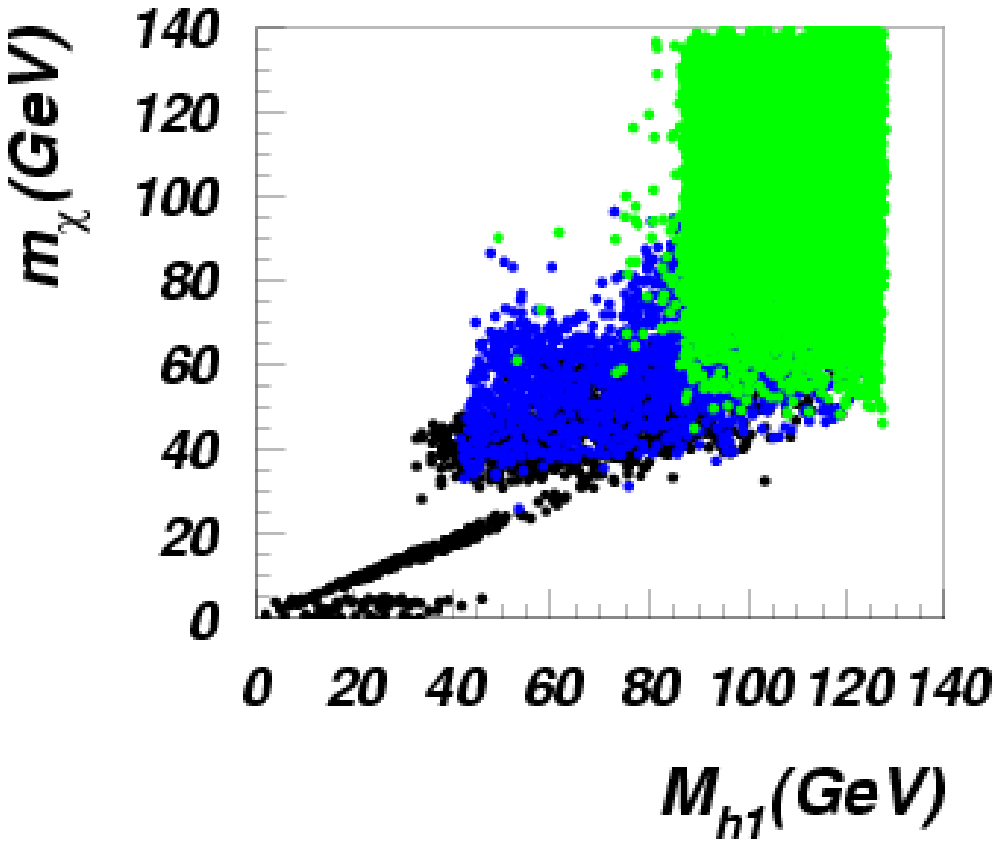}\hspace*{-0.9cm}
  \includegraphics[width=0.38\textwidth]{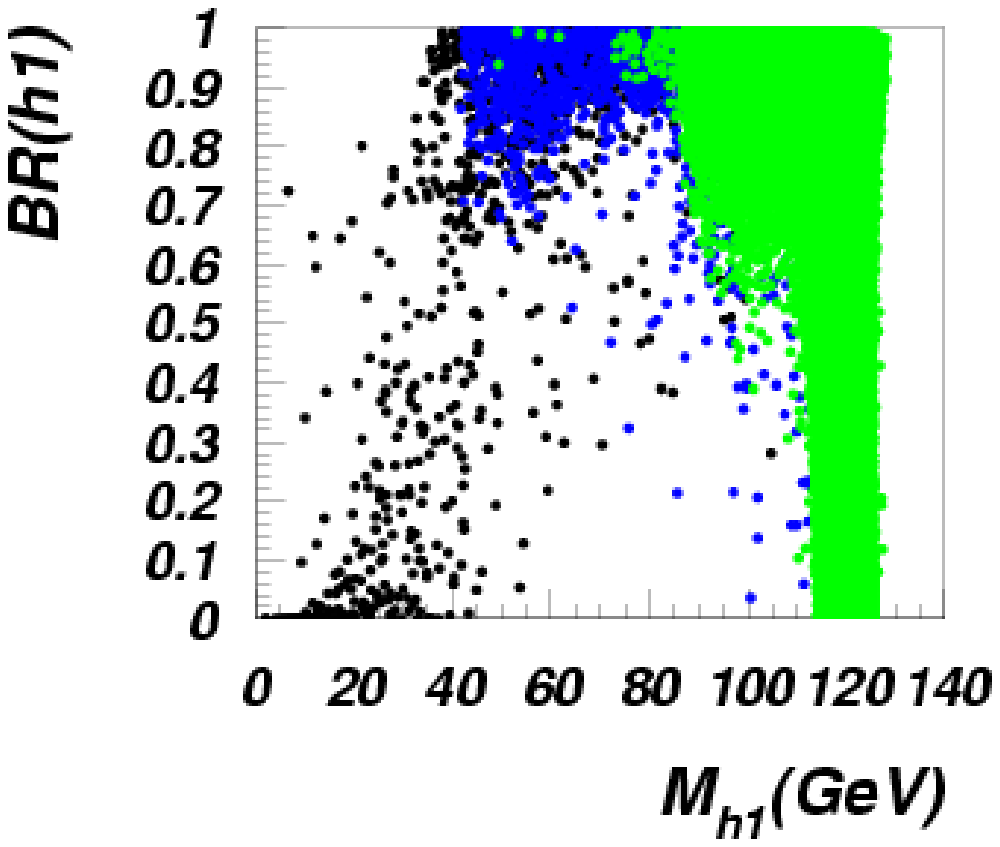}\hspace*{-0.9cm}
  \includegraphics[width=0.38\textwidth]{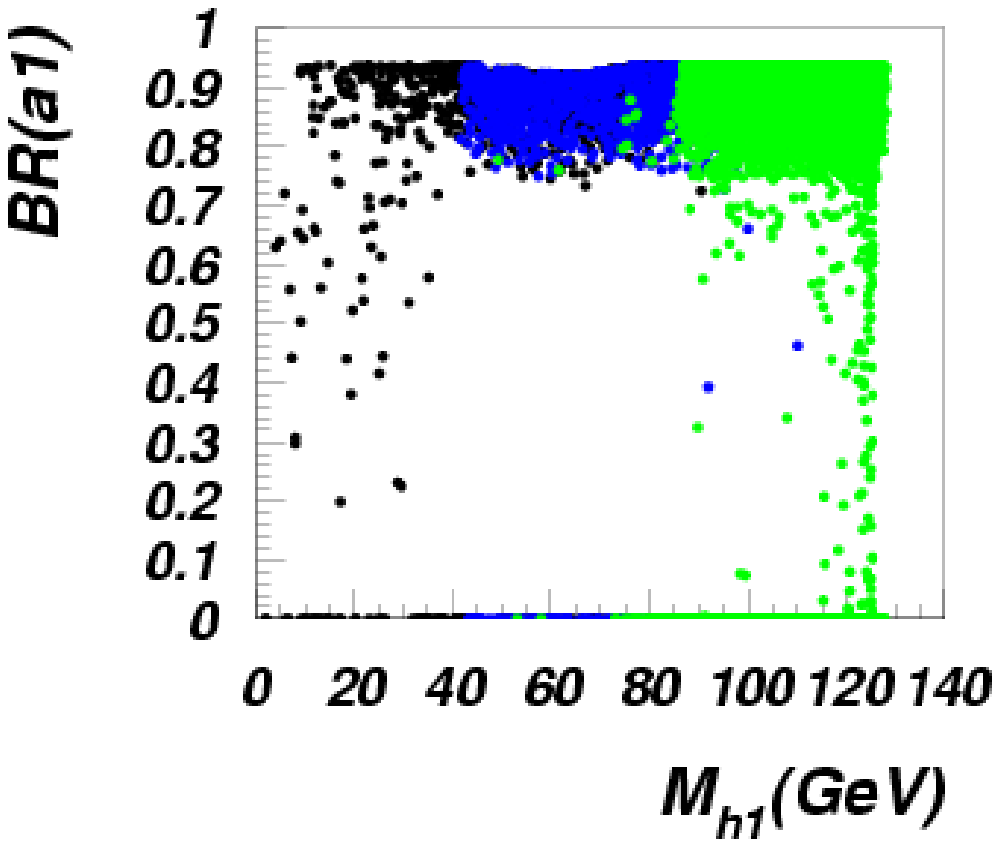}\hspace*{-0.9cm}
\vskip -2.5cm
\hspace*{0.33\textwidth}\hspace*{-5.5cm}{\bf (d)}
\hspace*{0.33\textwidth}\hspace*{-0.6cm}{\bf (e)}
\hspace*{0.33\textwidth}\hspace*{-0.6cm}{\bf (f)}
\vskip  2cm
\caption{\label{final-scan1}
Results of the `final' NMSSM parameter scan,
i.e., with $M_{a_1}< 10$~GeV and with Eq.~(\ref{cut:ak_narrow}) enforced:
(a) $R_{ZZh}$ (see text) versus $M_{h_1}$,
(b) $\kappa$ versus $A_\kappa$,
(c) $R_{ZZh}$ versus the singlino component $y_{15}^2$ of the lightest
neutralino $\chi^0_1$,
(d) $m_{\chi^0_1}$ versus $M_{h_1}$,
(e) BR($h_1 \to a_1 a_1$) versus $M_{h_1}$,
(f) BR($a_1 \to \tau^+\tau^-$) versus $M_{h_1}$.
The colour coding is the same as in Fig.~\ref{fig:scan-final}.}
\end{figure}

\item The branching ratios BR($h_1 \to a_1 a_1$)
  (Fig.~\ref{final-scan1}(e)) and  BR($a_1 \to \tau^+\tau^-$)
  (Fig.~\ref{final-scan1}(f)) are large for a large fraction of the
  parameter points with $M_{a_1}< 2m_b\approx10$ GeV which is
  crucial for the $h_1 \to a_1 a_1 \to 4 \tau$ mode analysed in the
  next section.

\end{enumerate}

\section{Phenomenology of the light $a_1$ scenario}

As a final step of our analysis, we have combined the production rates
of VBF and HS  
with selection efficiencies evaluated by generating these processes within the 
PYTHIA Monte Carlo (MC) generator. 
The latter have been estimated in presence of
experimental-like cuts, after parton showering and hadronisation and
with underlying events turned on.  
The jets, in particular in the VBF analysis, were found with the PYTHIA
routine PYCELL using  
the approximate CMS calorimeter tower granularity and cone size
0.5. No detector smearing  
of the jet energy is applied, however, we do not expect a 
significant reduction of the 
event selection efficiency once this is enforced\footnote{The selection efficiency 
is related to a cut on 
jet transverse energy $E_T$ and  since we  
cut right on the peak of the tagging jet $E_T$ distribution
we do not expect jet energy smearing to affect this efficiency substantially.}.

\begin{figure}[htb!]
\begin{center}
\includegraphics[width=.5\textwidth]{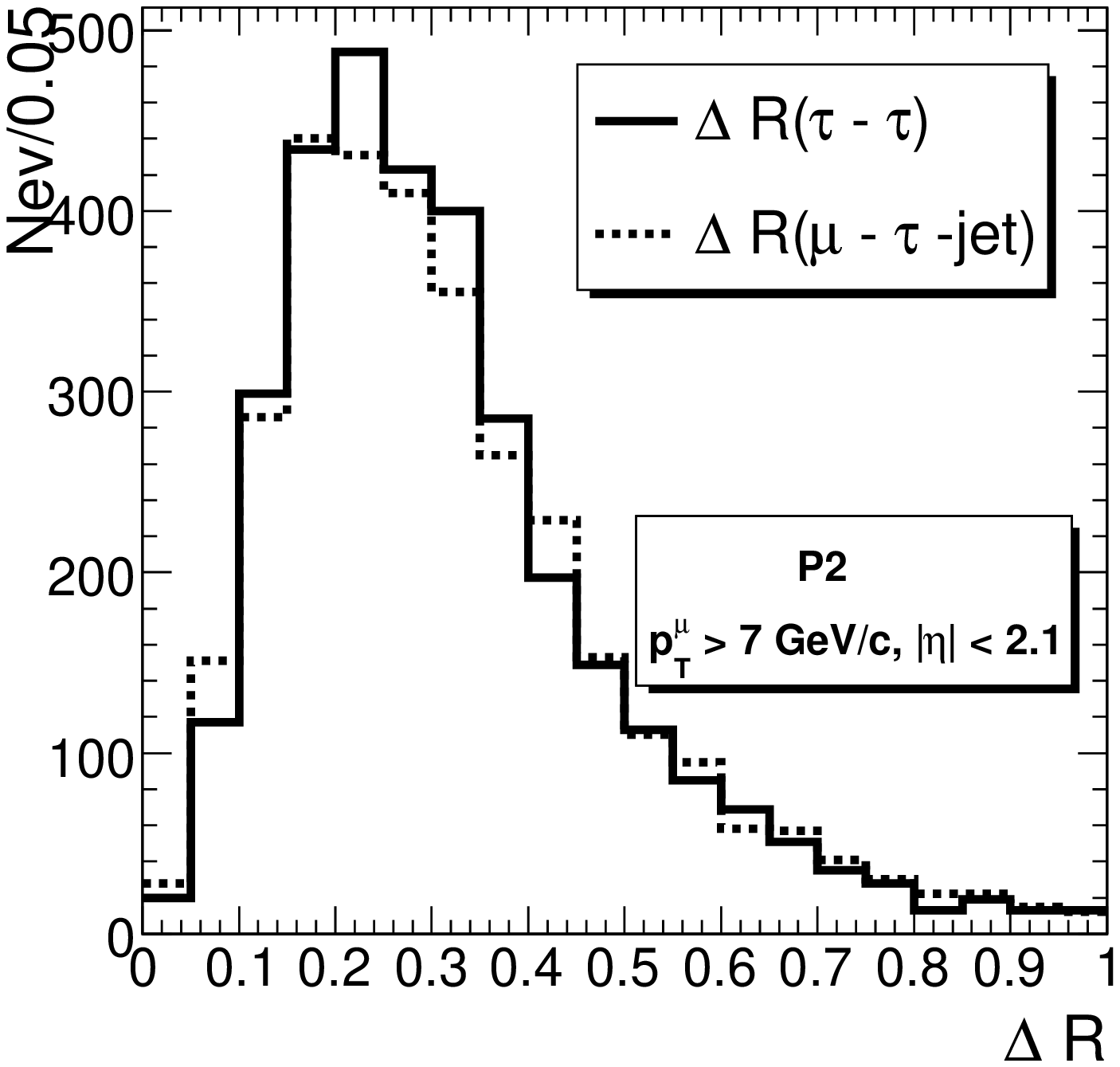}%
\includegraphics[width=.5\textwidth]{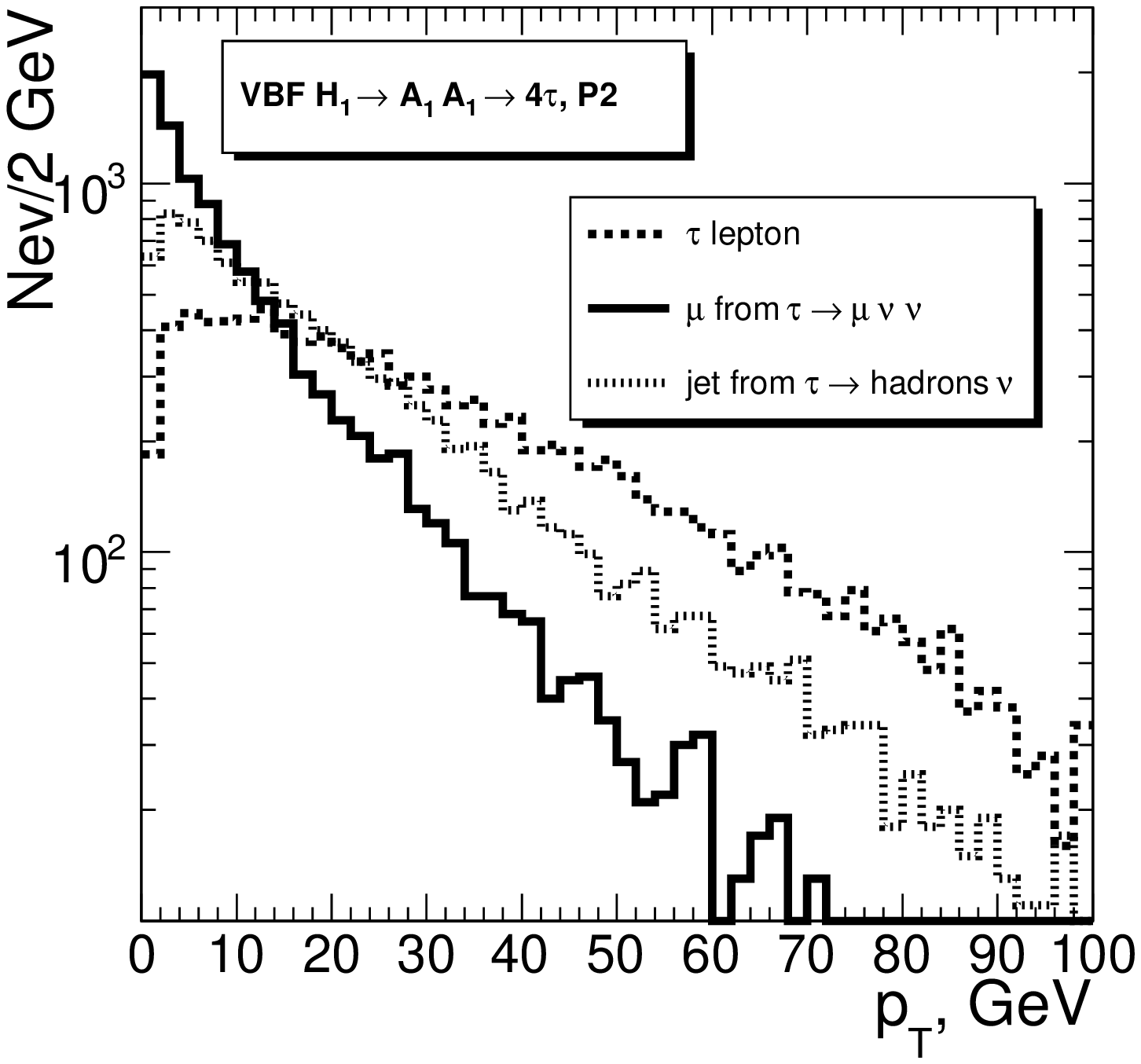}
\vskip -2.cm
\hspace*{0.5\textwidth}\hspace*{-15.1cm}{\bf (a)}
\hspace*{0.5\textwidth}\hspace*{-0.7cm}{\bf (b)}
\vskip  1cm
\caption{(a) Separation in ($\eta$, $\phi$) space between
         the two $\tau$ leptons and between the muon and the $\tau$ jet 
         from the $a_{1} \to \tau \tau \to \mu \nu \nu + \tau _{\rm jet} \nu $ 
         decay chain in the VBF
         $h_{1} \to a_1a_1\to \tau^+\tau^-\tau^+\tau^- \to 
         \mu^{\pm} \mu^{\pm} \tau _{\rm jet}^{\mp} \tau _{\rm jet}^{\mp}$ 
         channel for the benchmark point P2: 
         $M_{h_{1}}$=120.2 GeV, $M_{a_{1}}=9.1$ GeV.
         (b) Distributions of $p_{T}^{\tau}$, $p_{T}^{\mu}$ and
         $p_{T}^{\tau ~ \rm jet}$ for the point P2.}
\label{fig:cms_vbf1}
\end{center}
\end{figure}

The $h_{1} \to a_1a_1\to \tau^+\tau^-\tau^+\tau^-$ mode from VBF 
$h_{1}$ production has been considered for the 
$\mu^{\pm} \mu^{\pm} \tau _{\rm jet}^{\mp} \tau _{\rm jet}^{\mp}$ 
final state containing two same sign muons and two same sign $\tau$ jets
($\tau$ jet means the $\tau$ lepton decaying hadronically, 
$\tau \rightarrow \rm hadrons~+~\nu$). The $\tau$ leptons from the light 
$a_{1}$ decay are approximately collinear. Fig.~\ref{fig:cms_vbf1}(a) 
shows the separation in the ($\eta$, $\phi$) space between the two $\tau$
leptons and  between the muon and the $\tau$ jet in the decay chain 
$a_{1} \to \tau \tau \to \mu \nu \nu + \tau _{\rm jet} \nu$
for the benchmark point P2, $M_{h_{1}}=120.2$ GeV and $M_{a_{1}}=9.1$ GeV, 
proposed in \cite{Djouadi:2008uw,cNMSSM-benchmarks}.
The non-isolated, di-muon High Level (HL) trigger is needed to select the
signal events.
The standard CMS di-muon trigger with relaxed isolation has a di-muon
threshold 10 GeV on both muons for $\mathcal{L}=2\times$10$^{33}$~cm$^{-2}$s$^{-1}$
\cite{cms:2006}.
The muons from the signal events are very soft as one can see in 
Fig.~\ref{fig:cms_vbf1}(b), thus lower thresholds are needed. 
For example with the 7 GeV threshold the efficiency is increased approximately
by a factor of two. 
However the QCD background rate is also increased by approximately a factor of 
two \cite{cms:2002} which is not acceptable. 
In order to cope with the rate, the same sign relaxed di-muon trigger was 
introduced recently in the CMS trigger table \cite{cms:2007}. The
rate of di-muons from $b \bar{b}$ production is
reduced by a factor of four when asking for two muons of the 
same sign at the threshold 5 GeV. The off-line selection strategy
requires the presence of the two same sign, non-isolated muons with
one track within a cone 0.6 around each muon direction, thus selecting
the one prong $\tau$ decays. The full list of off-line selections is: 
\begin{itemize}
\item two same sign muons with $p_{T}>7$ GeV and $|\eta|<2.1$ with one 
      track of $p_{T}>2$ GeV in a cone 0.6 around each muon. The muon 
      and the track should have opposite charge;
\item two $\tau$ jets with $p_{T}>10$ GeV, $|\eta|<2.1$;
\item two jets with $p_{T}>30$ GeV, $|\eta|<4.5$.
\end{itemize}

The $h_{1} \to a_{1} a_{1} \to \tau^+\tau^-\tau^+\tau^-$ mode from HS 
production with leptonic decays of the $W$ bosons can give a very clean, almost
background free signal. The lepton coming from the $W$ decay is used for 
the triggering of the event using the CMS single isolated lepton trigger
with the thresholds 19 GeV and 26 GeV, respectively, for the muon and
electron \cite{cms:2002}.
The final state $\ell + \mu \tau _{\rm jet} \mu \tau _{\rm jet}$
($\ell=e$ or $\mu$, where the lepton comes from $W$ decay) has been
considered and
$\tau$ jets with one track have been selected. For each $\tau$-jet candidate 
there must be a muon in a cone 0.6 around the track. Unlike the VBF case, the
muons are not required to have the same sign. The full list of selections is:
\begin{itemize}
\item trigger selection: isolated muon or electron 
      with $p_{T}$ greater than 19 or 26 GeV, respectively, and
      $|\eta| <  2.5$;
\item two 1-prong $\tau$ jets with $p_{T}>10$ GeV, $|\eta|<2.1$. For each
      $\tau$ jet one muon of $p_{T}> 7$ GeV within a cone 0.6 around
      the $\tau$ 
      track should exist and have a charge opposite to the track charge;
\item events with extra jets in addition to the two $\tau$ jets are rejected.
\end{itemize}

\begin{figure}[!t]
  \includegraphics[width=0.55\textwidth]{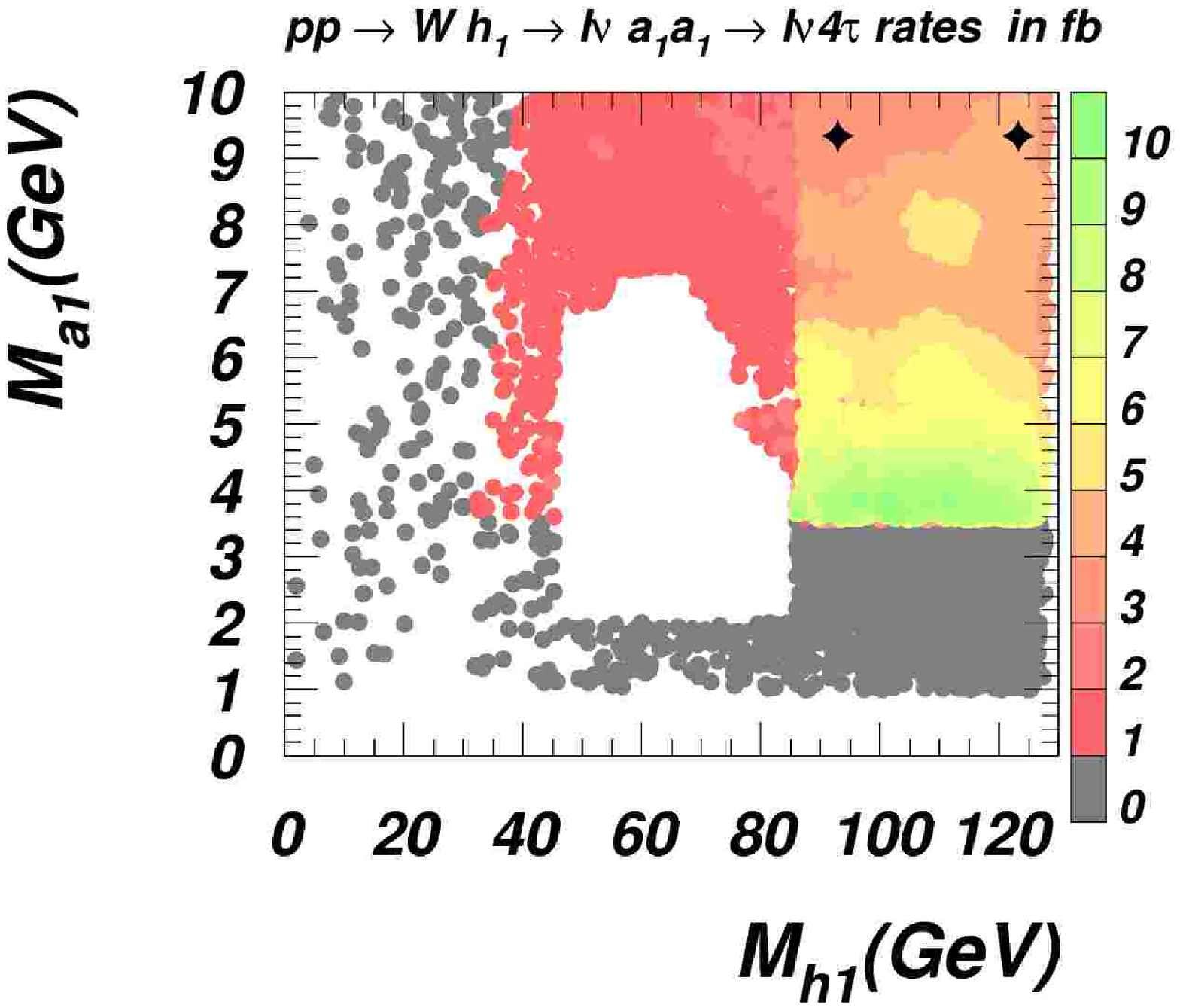}%
  \hspace*{-0.8cm}
  \includegraphics[width=0.55\textwidth]{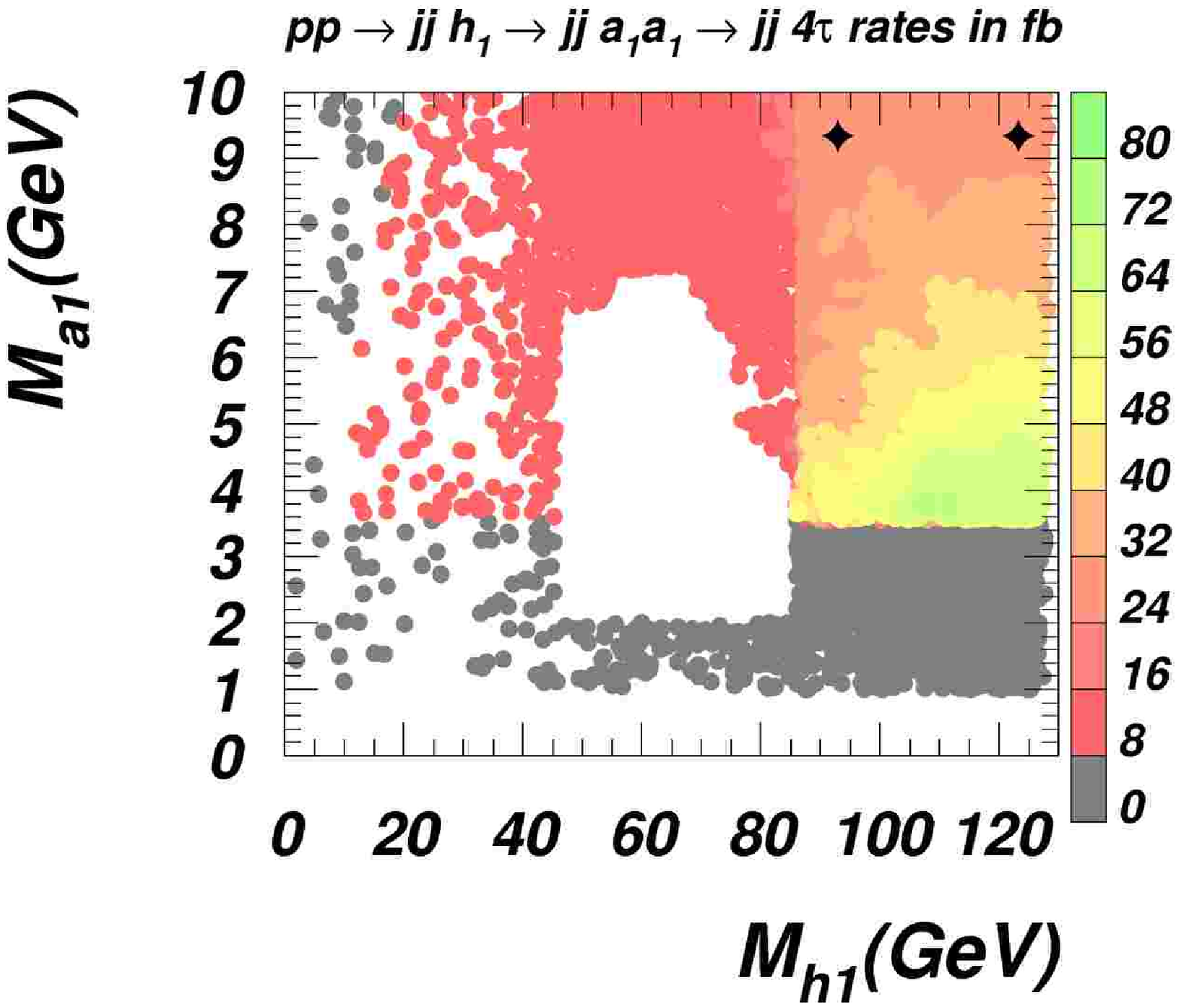}%
\vskip -3.cm
\hspace*{0.55\textwidth}\hspace*{-8cm}{\bf (a)}
\hspace*{0.55\textwidth}\hspace*{-1.5cm}{\bf (b)}
\vskip  2cm
  \caption{\label{final-rates}
Cross sections (including all relevant branching ratios) for HS
(a) and VBF (b) after the selection cuts described in the text. 
The population of points used correspond
to that of the `final scan' described previously.
Black diamonds correspond to the benchmark P2 (right) and P3 (left) from Refs.~\cite{Djouadi:2008uw,cNMSSM-benchmarks}.}
\end{figure}

The results in Fig.~\ref{final-rates} show that, after our final scan, the 
population of parameter points is such that in both channels the highest
cross sections are found for {$M_{h_1}\gtrsim M_Z$, although
in the case of VBF also lower $h_1$ masses can yield sizable rates. }
Independently
of $M_{h_1}$, the $a_1$ mass  enables sizable event rates anywhere above
$2m_\tau$, but particularly just above the threshold. 
At high luminosity, 100 fb$^{-1}$, the highest rates would correspond to
1000 events per year for HS and 8000 for VBF. 
The black diamond symbols in Fig.~\ref{final-rates} denote the two
NMSSM benchmarks points P2 (right point) and P3 (left point), defined in
\cite{Djouadi:2008uw,cNMSSM-benchmarks}.
As it can be appreciated, they correspond to event rates that are mid
range amongst all those explored, hence not particularly biased 
towards a far too favourable NMSSM setup, yet susceptible to
experimental discovery.

\section{Conclusions}

We have analysed the CP-even Higgs boson $h_1$ in the NMSSM decaying
into $a_1a_1$ pairs in turn yielding four $\tau$ leptons, produced at
the LHC in HS and VBF production and searched for through their
semi-leptonic/hadronic decays into muons and jets. We have found 
that there is significant potential, especially via the VBF production
channel, to discover a Higgs boson at the LHC in NMSSM scenarios with
$M_{a_1}<2m_b$ which is an important step 
in establishing a no-lose theorem for the NMSSM.
In order to achieve this the same sign di-muon trigger with lower
threshold \cite{cms:2007} is crucial.
We have restricted ourselves to the case $M_{a_1}<2m_b$,
{where the $h_1$ decay fraction into $\tau$'s
in enhanced }
(otherwise $a_1\to b\bar b$ decays are dominant).
A scan of the low-energy NMSSM parameter space without assuming
unification at the high scale has shown 
that the $h_1$ state can be very light, indeed at times lighter than the 
${a_1}$.  {This configuration does not take place in the constrained
NMSSM (cNMSSM) of Ref. \cite{Djouadi:2008uw,Djouadi:2008yj}.
However, we are currently investigating whether such light $h_1$ masses
can be found in the NMSSM with non-universal boundary conditions at the unification scale
\cite{future}.}
With  reference to the NMSSM benchmarks points defined in
\cite{Djouadi:2008uw,cNMSSM-benchmarks}, we note
that those relevant to our $4\tau$ channel 
{are the P2 and P3 benchmarks,
which yield event rates in the mid range amongst those explored here}.
Our summary is preliminary, as only signal processes have been
considered and only in presence of MC simulations, 
with no backgrounds and full detector performance enabled. The latter
clearly ought to be investigated before drawing 
any firm conclusions and this is currently being done.
{However, we would like to conclude that the $2\mu+2$jets signature 
from $h_1\to a_1a_1\to 4\tau$ decays with the $h_1$ produced in HS and
VBF processes does produce sizable event rates 
as high as 1000 events per year for HS and 8000 for VBF
so that 
 we expect that a significant part of the viable NMSSM parameter space
 will be covered by using the signature and selection
cuts suggested in our study.}

\end{document}